# Correlation between amygdala BOLD activity and frontal EEG asymmetry during real-time fMRI neurofeedback training in patients with depression

Vadim Zotev[1#], Han Yuan[1], Masaya Misaki[1], Raquel Phillips[1], Kymberly D. Young[1], Matthew T. Feldner[2], and Jerzy Bodurka[1,3,4#]

[1]Laureate Institute for Brain Research, Tulsa, OK, USA; [2]Department of Psychological Science, University of Arkansas, Fayetteville, AR, USA
[3]Center for Biomedical Engineering, University of Oklahoma, Norman; [4]College of Engineering, University of Oklahoma, Norman, OK, USA

**Abstract:** Real-time fMRI neurofeedback (rtfMRI-nf) is an emerging approach for studies and novel treatments of major depressive disorder (MDD). EEG performed simultaneously with an rtfMRI-nf procedure allows an independent evaluation of rtfMRI-nf brain modulation effects. Frontal EEG asymmetry in the alpha band is a widely used measure of emotion and motivation that shows profound changes in depression. However, it has never been directly related to simultaneously acquired fMRI data. We report the first study investigating electrophysiological correlates of the rtfMRI-nf procedure, by combining the rtfMRI-nf with simultaneous and passive EEG recordings. In this pilot study, MDD patients in the experimental group ($n$=13) learned to upregulate BOLD activity of the left amygdala using an rtfMRI-nf during a happy emotion induction task. MDD patients in the control group ($n$=11) were provided with a sham rtfMRI-nf. Correlations between frontal EEG asymmetry in the upper alpha band and BOLD activity across the brain were examined. Average individual changes in frontal EEG asymmetry during the rtfMRI-nf task for the experimental group showed a significant positive correlation with the MDD patients' depression severity ratings, consistent with an inverse correlation between the depression severity and frontal EEG asymmetry at rest. The average asymmetry changes also significantly correlated with the amygdala BOLD laterality. Temporal correlations between frontal EEG asymmetry and BOLD activity were significantly enhanced, during the rtfMRI-nf task, for the amygdala and many regions associated with emotion regulation. Our findings demonstrate an important link between amygdala BOLD activity and frontal EEG asymmetry during emotion regulation. Our EEG asymmetry results indicate that the rtfMRI-nf training targeting the amygdala is beneficial to MDD patients. They further suggest that EEG-nf based on frontal EEG asymmetry in the alpha band would be compatible with the amygdala-based rtfMRI-nf. Combination of the two could enhance emotion regulation training and benefit MDD patients.

## 1. Introduction

Major depressive disorder (MDD) is characterized by functional impairments affecting prefrontal, limbic, striatal, thalamic, and basal forebrain structures (Price & Drevets, 2010). Common treatments for MDD include cognitive behavioral therapy (CBT), antidepressant medication therapy, and the combination of the two (Driessen & Hollon, 2010). Unfortunately, only 35-55% of MDD patients undergoing CBT achieve remission (DeRubeis et al., 2005; Dimidjian et al., 2006). Recent years have seen a growing interest in real-time fMRI neurofeedback (rtfMRI-nf) as a potential tool for studies and treatment of neuropsychiatric disorders. rtfMRI-nf enables volitional regulation of blood-oxygenation-level-dependent (BOLD) activity of target brain regions in real time (for reviews, see deCharms, 2008; Sulzer et al., 2013; Weiskopf, 2012). This approach is non-invasive, spatially precise, and capable of targeting deep brain structures such as the amygdala.

Several pilot studies have explored the feasibility of emotion regulation training with rtfMRI-nf in patients with neuropsychiatric disorders. They included self-regulation of the anterior insula (Caria et al., 2007, 2010) in patients with schizophrenia (Ruiz et al., 2013), self-regulation of functionally localized emotional networks (Johnston et al., 2010, 2011) in patients with MDD (Linden et al., 2012), and self-regulation of the left amygdala (Zotev et al., 2011, 2013a) in MDD patients (Young et al., 2014). These proof-of-concept studies each reported success in rtfMRI-nf training and improvements in the patients' mental states.

Advances in simultaneous EEG-fMRI technique (e.g. Mulert & Lemieux, 2010) have made it possible to perform an rtfMRI-nf procedure with simultaneous EEG recordings, and even provide simultaneous multimodal rtfMRI and EEG neurofeedback (rtfMRI-EEG-nf) (Zotev et al., 2014). The combination of rtfMRI-nf and simultaneous (passive) EEG acquisition offers new important opportunities for research and neurotherapy applications of rtfMRI-nf in depression. *First*, electrophysiological

[#]Corresponding authors. E-mail: vzotev@laureateinstitute.org (V. Zotev), jbodurka@laureateinstitute.org (J. Bodurka)

correlates of rtfMRI-nf training can be identified and evaluated based on the broad existing knowledge of brain electrophysiology in depression. *Second*, relationships between BOLD activities of brain regions targeted by rtfMRI-nf (such as the amygdala) and electrophysiological measures relevant to depression can be elucidated. *Third*, target EEG measures can be identified and used to implement either the rtfMRI-EEG-nf (Zotev et al., 2014) or EEG-only neurofeedback (EEG-nf) (e.g. Gruzelier, 2014) for more efficient and/or more portable neurotherapies for depression. Notably, no rtfMRI-nf studies with simultaneous EEG recordings have been reported other than our proof-of-concept work on the multimodal rtfMRI-EEG-nf (Zotev et al., 2014).

Numerous EEG studies of human emotion and motivation have examined frontal EEG asymmetry (for reviews, see Coan & Allen, 2004; Davidson, 1992, 1998; Harmon-Jones et al., 2010). Frontal EEG asymmetry, which we abbreviate as FEA, is commonly defined for the alpha EEG band as $\ln(P(\text{right})) - \ln(P(\text{left}))$, where $P$ is the alpha power for corresponding frontal EEG channels on the right and on the left. The FEA reflects functional differences between the approach and avoidance motivation systems (e.g. Elliot & Covington, 2001). The approach-withdrawal hypothesis (e.g. Davidson, 1998; Tomarken & Keener, 1998) posits that the approach motivation system recruits activity of the left prefrontal regions, leading to reduced alpha EEG power on the left and more positive FEA, while the avoidance motivation system engages activity of the right prefrontal regions, leading to reduced alpha power on the right and more negative FEA values (see also De Pascalis et al., 2013; Pizzagalli et al., 2005; Sutton & Davidson, 1997). Frontal asymmetries associated with emotion/motivation have also been observed for the theta EEG band (e.g. Aftanas & Golocheikine, 2001; Ertl et al., 2013) and the high-beta EEG band (e.g. Paquette et al., 2009; Pizzagalli et al., 2002).

The main EEG findings regarding the approach-avoidance lateralization have been confirmed by independent fMRI studies (Berkman & Lieberman, 2010; Canli et al., 1998; Herrington et al., 2005, 2010; Spielberg et al., 2011, 2012). These studies more specifically associated the approach and avoidance motivation systems with the left and right dorsolateral prefrontal cortex (DLPFC), respectively. fMRI studies also demonstrated that the amygdala plays an important role in both the approach and avoidance motivation systems (e.g. Cunningham et al., 2005, 2010; Schlund & Cataldo, 2010; Spielberg et al., 2012). In particular, the motivational salience hypothesis posits that the amygdala activity is closely linked to motivational relevance of stimuli (Cunningham et al., 2010).

MDD patients consistently exhibit significantly lower FEA values at rest than healthy individuals (e.g. Gotlib et al., 1998; Henriques & Davidson, 1991; Keune et al., 2013; Stewart et al., 2011; Thibodeau et al., 2006). This phenomenon is associated with hypoactivity of the left prefrontal regions (Henriques & Davidson, 1991), which indicates deficient approach motivation in depressed individuals, leading to their diminished reward sensitivity and ability to experience pleasure (i.e. anhedonia). Importantly, the FEA reflects both emotional traits, such as vulnerability to depression, and emotional states (Coan & Allen, 2004). The FEA is more positive for approach-related emotions (such as happiness), and more negative for avoidance-related emotional states (such as fear) (Coan et al., 2001; Davidson et al., 1990). Positive FEA changes can be achieved through positive emotion induction, mindfulness meditation (e.g. Keune et al., 2013), as well as explicit FEA manipulation by means of EEG-nf. Several EEG-nf studies of emotion regulation have used the FEA as a target measure (Allen et al., 2001; Baehr et al., 1997; Cavazza et al., 2014; Choi et al., 2011; Peeters et al., 2014a, 2014b; Rosenfeld et al., 1995), or led to significant changes in resting FEA (e.g. Paquette et al., 2009).

Remarkably, despite the facts that the FEA in the alpha band has been used as a measure of emotion and motivation in hundreds of EEG studies, and the FEA abnormalities have been commonly reported in MDD, the FEA has never been directly related to simultaneously acquired fMRI data (see Sec. 4).

Because an rtfMRI-nf training in general is a volitional regulation of one's brain activity toward a certain goal (i.e. goal pursuit), motivation plays an important role. A stronger approach motivation can conceivably lead to a better performance of an rtfMRI-nf task, while a stronger avoidance motivation can impair the performance. Therefore, the FEA is a relevant measure for evaluation of rtfMRI-nf effects. It may be particularly useful in the case of an rtfMRI-nf of the amygdala. Because the amygdala is a part of both the approach and avoidance motivation systems, as mentioned above, regulation of the amygdala BOLD activity by means of rtfMRI-nf should be accompanied by modulation of these systems, leading to modulation of the FEA.

Here we report the first and well controlled pilot study in which EEG recordings, performed simultaneously with rtfMRI-nf training, were used to evaluate electrophysiological effects of the rtfMRI-nf. In this work, MDD patients learned to upregulate BOLD activity of their left amygdala using rtfMRI-nf during a happy emotion induction task. We chose the amygdala as a target for rtfMRI-nf, because the amygdala activity shows profound changes in MDD (Price & Drevets, 2010), including blunted activation in response to positive emotional stimuli (Murray et al., 2011). We employed the same rtfMRI-nf paradigm as in our previous studies with healthy participants (Zotev et al., 2011) and MDD patients (Young et al., 2014). We aimed to investigate EEG correlates of this paradigm to better understand its mechanisms and effects in MDD.

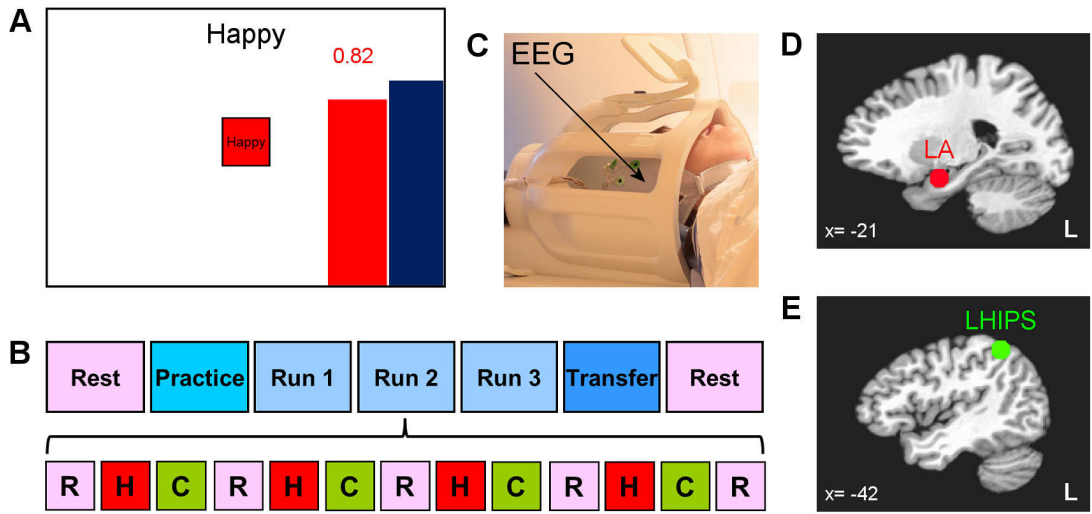

Figure 1. **Experimental paradigm for real-time fMRI neurofeedback training of emotional self-regulation with simultaneous EEG. A)** Real-time display screen for Happy Memories conditions with real-time fMRI neurofeedback (rtfMRI-nf). The variable-height rtfMRI-nf bar is red, and the target level bar is blue. **B)** Protocol for the rtfMRI-nf experiment included seven runs, each lasting 8 min 46 s: Rest (RE), Practice (PR), Run 1 (R1), Run 2 (R2), Run 3 (R3), Transfer (TR), and Rest (RE). The experimental runs (except the Rest) consisted of 40-s long blocks of Happy Memories (H), Count (C), and Rest (R) conditions. **C)** An MR-compatible 32-channel EEG system was used to perform EEG recordings simultaneously with fMRI data acquisition. **D)** Target region of interest (ROI) in the left amygdala (LA) region for the experimental group (EG). **E)** Target ROI in the left horizontal segment of the intraparietal sulcus (LHIPS) region for the control group (CG).

We used our EEG-fMRI data acquired in this study to test two main hypotheses. *First*, we hypothesized that the participants receiving the amygdala-based rtfMRI-nf would show positive FEA changes during the rtfMRI-nf task, indicating an enhancement in approach motivation, compared to control participants receiving a sham rtfMRI-nf. We also expected to observe some dependence of the FEA changes on the MDD patients' depression severity. *Second*, we hypothesized that FEA variations during the rtfMRI-nf task targeting the amygdala would exhibit a temporal correlation with the simultaneously measured BOLD activity of the amygdala.

## 2. Methods

### 2.1. Participants

The study was performed at the Laureate Institute for Brain Research, and was approved by the Western Institutional Review Board. All study procedures were conducted in accordance with the principles expressed in the Declaration of Helsinki.

Twenty four unmedicated MDD patients completed two sessions of the emotion self-regulation study involving rtfMRI-nf training. In the first session, neurofeedback-naive MDD patients learned to upregulate BOLD activity of the amygdala using rtfMRI-nf while performing a happy emotion induction task. Results of this session have been reported previously (Young et al., 2014). In the second session, the same MDD patients followed the same procedure, except that they had to wear an MR-compatible EEG cap and EEG recordings were performed simultaneously with fMRI. Here we report results for this second rtfMRI-nf session.

All the participants met the Diagnostic and Statistical Manual of Mental Disorders (DSM-IV-TR) (American Psychiatric Association, 2000) criteria for MDD in a current major depressive episode. Prior to the rtfMRI-nf session, the participants underwent a psychological evaluation consisting of multiple well-established measures of depression and related features, administered by a licensed psychiatrist. The evaluation included the 21-item Hamilton Depression Rating Scale (HDRS, Hamilton, 1960), the Montgomery-Asberg Depression Rating Scale (MADRS, Montgomery & Asberg, 1979), the Hamilton Anxiety Rating Scale (HARS, Hamilton, 1959), the Snaith-Hamilton Pleasure Scale (SHAPS, Snaith et al., 1995), and the 20-item Toronto Alexithymia Scale (TAS-20, Bagby et al., 1994). Both before and after the rtfMRI-nf session, the participants completed the Profile of Mood States (POMS, McNair et al., 1971), the State-Trait Anxiety Inventory (STAI, Spielberger et al., 1970), and the Visual Analogue Scale (VAS) with 10-point subscales for happy, restless, sad, anxious, irritated, drowsy, and alert states.

Participants in the experimental group (EG, $n=13$, 9 females) received rtfMRI-nf based on BOLD activity of the left amygdala (LA) target region (Zotev et al., 2011). Participants in the control group (CG, $n=11$, 9 females) were provided, without their knowledge, with sham rtfMRI-nf based on BOLD activity of the left horizontal segment of the intraparietal sulcus (LHIPS) region, presumably not involved in emotion processing (Zotev et al., 2011). (Compared to the initial report by Young et al., 2014, four more MDD patients completed both sessions in the CG group, and one of the EG participants did not finish the second session). The participants' average age was 41 ($SD=9$) years for the EG and 34 ($SD=8$) years for the CG. The groups' age difference was not significant ($t(22)=1.88$, $p<0.073$). All the participants were right-handed. The psychological trait measures for the EG and CG participants before the second rtfMRI-nf session are reported in *Supplementary material* (Table S1). Importantly, the two groups did not differ in their mean depression, anxiety, anhedonia, and alexithymia ratings (Table S1).

*2.2. Experimental paradigm*

The experimental paradigm (Fig. 1) was developed based on the results of our previous rtfMRI-nf study with healthy participants (Zotev et al., 2011). The rtfMRI-nf signal was presented to a subject inside the MRI scanner as a variable-height red bar on the screen (Fig. 1A). The bar height represented BOLD activity (fMRI percent signal change with respect to a resting baseline) in a target region of interest (ROI). The target ROIs were defined as 14-mm diameter spheres in the Talairach space (Talairach & Tournoux, 1988) as described previously (Zotev et al., 2011). They were centered, respectively, at $(-21, -5, -16)$ in the LA region for the EG (Fig. 1D), and at $(-42, -48, 48)$ in the LHIPS region for the CG (Fig. 1E). The specified ROI centers were selected based on quantitative meta-analyses of functional neuroimaging studies investigating the role of the amygdala in emotion processing (Sergerie et al., 2008) or the role of the HIPS in number processing (Dehaene et al., 2003). The height of the red rtfMRI-nf bar was updated every 2 s. The height of the blue target bar was adjusted incrementally between runs.

The rtfMRI-nf training protocol (Fig. 1B) included three conditions: Happy Memories, Count, and Rest. For the Happy Memories conditions with rtfMRI-nf, the participants were instructed to feel happy by evoking and contemplating happy autobiographical memories while attempting to simultaneously raise the level of the red rtfMRI-nf bar toward the fixed level of the blue target bar (Fig. 1A). For the Count conditions, the subjects were asked to mentally count back from 300. For the Rest conditions, the participants were asked to relax while looking at the screen. No bars were displayed during the Count and Rest conditions.

The rtfMRI-nf experiment included seven runs, and each run (except the two Rest runs) consisted of alternating 40-s blocks of Happy Memories, Count, and Rest conditions (Fig. 1B). The target level for the rtfMRI-nf (the blue bar height) was set to 0.5%, 1.0%, 1.5%, and 2.0% for the Practice run, Run 1, Run 2, and Run 3, respectively. During the Transfer run, the participants performed the same emotion induction task, but no bars were shown for the Happy Memories conditions. The Count condition involved counting back from 300 by subtracting 3, 4, 6, 7, and 9 for the Practice run, Run 1, Run 2, Run 3, and the Transfer run, respectively. After each experimental run with the Happy Memories task, the participants were asked to verbally rate their performance on a scale from 0 (not at all) to 10 (extremely) by answering two questions: "How successful were you at recalling your happy memories?" and "How happy are you right now?". All details of the experimental paradigm have been described previously (Young et al., 2014).

*2.3. Data acquisition*

All experiments were conducted on the General Electric Discovery MR750 3T MRI scanner with a standard 8-channel receive-only head coil (Fig. 1C). A single-shot gradient echo EPI sequence with FOV/slice=240/2.9 mm, $TR/TE$=2000/30 ms, flip angle=90°, 34 axial slices per volume, slice gap=0.5 mm, SENSE $R$=2 in the phase encoding (anterior-posterior) direction, acquisition matrix 96×96, sampling bandwidth=250 kHz, was employed for fMRI. Each fMRI run lasted 8 min 46 s and included 263 EPI volumes (the first three EPI volumes were included to allow fMRI signal to reach a steady state and were excluded from data analysis). Physiological pulse oximetry and respiration waveforms were recorded simultaneously with fMRI. The EPI images were reconstructed into a 128×128 matrix, resulting in 1.875×1.875×2.9 mm$^3$ fMRI voxels. A T1-weighted 3D MPRAGE sequence with FOV/slice=240/1.2 mm, $TR$/$TE$=5.0/1.9 ms, $TD$/$TI$=1400/725 ms, flip angle=10°, 128 axial slices per slab, SENSE $R$=2, acquisition matrix 256×256, sampling bandwidth=31.2 kHz, scan time=4 min 58 s, was used for anatomical imaging. It provided structural brain images with 0.9375×0.9375×1.2 mm$^3$ voxels.

EEG recordings were performed simultaneously with fMRI (Fig. 1C) using an MR-compatible 32-channel EEG system from Brain Products GmbH. The EEG data were acquired with 0.2 ms temporal and 0.1 µV measurement resolution in 0.016...250 Hz frequency band with respect to FCz reference. All technical details of the EEG-fMRI system configuration and data acquisition have been described elsewhere (Zotev et al., 2012).

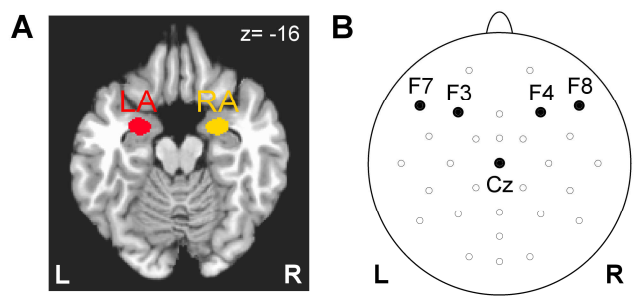

**Figure 2. Regions of interest and EEG channels used in offline analysis of amygdala BOLD activity and frontal EEG asymmetry. A)** Left amygdala (LA) and right amygdala (RA) ROIs were defined anatomically according to the co-planar stereotaxic atlas of the human brain by Talairach and Tournoux. The ROIs are projected in the figure onto the standard TT_N27 template in the Talairach space. The left hemisphere (L) is to the reader's left. **B)** Pairs of frontal EEG channels used to quantify frontal EEG asymmetry: F3 (left) and F4 (right), F7 (left) and F8 (right). Cz channel was used as a reference.

*2.4. Real-time data processing*

The rtfMRI-nf was implemented using the custom real-time fMRI system (Bodurka & Bandettini, 2008) utilizing real-time functionality of AFNI (Cox, 1996; Cox & Hyde, 1997) as described previously (Zotev et al., 2011). A high-resolution MPRAGE structural brain image and a short EPI dataset (5 volumes) were acquired prior to each rtfMRI-nf experiment. The last volume in the EPI dataset was used as a reference EPI volume defining the subject's individual EPI space. The LA and LHIPS target ROIs, defined in the Talairach space (Figs. 1D,E) were transformed to the individual EPI space using the MPRAGE image data. The resulting ROIs in the EPI space contained approximately 140 voxels each. During the subsequent fMRI runs (Fig. 1B), the AFNI real-time plugin was used to perform volume registration of each acquired EPI volume to the reference EPI volume (motion correction) and export mean values of fMRI signals for these ROIs in real time. The rtfMRI signal for each Happy Memories condition was computed as a percent signal change relative to the baseline obtained by averaging the fMRI signal for the preceding Rest condition block (Fig. 1B). Such block-specific baseline computation reduced effects of drifts in the fMRI data. A moving average of the current and two preceding rtfMRI signal values was computed to reduce effects of fMRI noise and physiological artifacts (Zotev et al., 2011). This average value was used to set the height of the red rtfMRI-nf bar (Fig. 1A) every $TR$=2 s.

*2.5. fMRI data analysis*

Offline analysis of the fMRI data was performed in AFNI as described in detail in *Supplementary material* (S1.1). The analysis involved fMRI pre-processing with cardiorespiratory artifact correction (Glover et al., 2000), slice timing correction and volume registration. fMRI signal-to-noise performance is illustrated in *Supplementary material* (Fig. S1). A general linear model (GLM) analysis with Happy Memories and Count block-stimulus conditions was applied to the preprocessed fMRI data. Average GLM-based fMRI percent signal changes were computed for the LA and RA ROIs, shown in Fig. 2A. The ROIs were defined anatomically as the amygdala regions specified in the AFNI implementation of the Talairach-Tournoux brain atlas. Such ROI definition has the advantage of being independent of any functional information. To compare BOLD activity levels for the LA and RA, we computed amygdala BOLD laterality, i.e. differences in the fMRI percent signal changes between the LA and RA ROIs for each task, run, and participant, as discussed in *Supplementary material* (S1.2). Similar fMRI laterality measures had been used before (e.g. Koush et al., 2013; Robineau et al., 2014).

*2.6. EEG data analysis*

Offline analysis of the EEG data, acquired simultaneously with fMRI, was performed using BrainVision Analyzer 2 software as described in detail in *Supplementary material* (S1.3). Removal of EEG artifacts was based on the average artifact subtraction and independent component analysis (Bell & Sejnowski, 1995; McMenamin et al., 2010). Channel Cz was selected as a new reference, because it had been most commonly used as a reference in FEA studies (Hagemann et al., 2001; Thibodeau et al., 2006). The reference selection is further discussed in *Supplementary material* (S1.3). Following the artifact removal, a continuous wavelet transform was applied to obtain EEG signal power for each channel in [0.25...15] Hz frequency range with 0.25 Hz frequency resolution and 8 ms temporal resolution. EEG power $P(t)$ was transformed toward a normal distribution using the $\ln(P(t))$ normalizing function. An EEG coherence analysis was performed for the Happy Memories conditions in each of the four rtfMRI-nf runs. It was conducted for all channel pairs with 0.244 Hz frequency resolution as described in *Supplementary material* (S1.3).

*2.7. Frontal EEG asymmetry analysis*

Time-dependent FEA was computed as $\ln(P(F4))-\ln(P(F3))$, where $P$ is EEG power as a function of time in the upper alpha EEG band for a given channel (F3 or F4). The FEA for channels F7 and F8, which is also relevant to depression (Thibodeau et al., 2006), was defined in a similar way. These four channels are depicted schematically in Fig. 2B. We focused on the upper alpha EEG band, because resting upper alpha (alpha2) EEG activity of the left prefrontal regions had been shown to exhibit a significant inverse correlation with individual reward responsiveness, a measure of approach motivation (Pizzagalli et al., 2005).

The upper alpha band was defined for each participant as [IAF…IAF+2] Hz, where IAF is an individual alpha peak frequency. The IAF was determined by inspection of average EEG power spectra for occipital and parietal channels for Rest condition blocks in the four rtfMRI-nf runs (Fig. 1B). In addition to the FEA, a power-sum function $\ln(P(F4))+\ln(P(F3))$ was computed for the same pair of channels and used to define covariates of no interest.

*2.8. EEG-fMRI correlation analysis*

To study task-specific temporal correlations between the FEA and BOLD activity, we performed a psychophysiological interaction (PPI) analysis (Friston et al., 1997) of the EEG-fMRI data. We tested the hypothesis that temporal correlations between the FEA and BOLD activity of brain regions involved in emotion/motivation would be stronger during the rtfMRI-nf task than during the control task (for the EG). This hypothesis includes two assumptions. *First*, it is assumed that performance of the rtfMRI-nf task targeting the LA engages and modulates activities of the motivation systems (in a way that is relevant to emotion regulation, more specifically – happy emotion induction). *Second*, it is assumed that the FEA is a meaningful measure of the resulting motivation systems' activity, as suggested by previous FEA studies.

The FEA computed with 8 ms temporal resolution was used to define two fMRI regressors for the PPI analysis. One regressor was obtained by convolution of the FEA (converted to *z*-scores) with the hemodynamic response function (HRF, 8 ms resolution). For the other regressor, the FEA (converted to *z*-scores) was first multiplied by the contrast function (equal to +1 for Happy Memories, –1 for Count, and 0 for Rest conditions), and then convolved with the HRF. Both regressors were sub-sampled to middle time points of fMRI volumes, linearly detrended, and included in the PPI analysis within the GLM framework. One PPI term ('correlation') described average correlation of the FEA-based regressor with the fMRI data across all three experimental conditions. The other PPI term ('interaction') described [FEA-based regressor] × [Happy–Count] interaction, which corresponded to the difference in correlations of the fMRI data and the FEA-based regressor between the Happy Memories and Count conditions. In addition to the FEA-based regressors, we defined a similar pair of regressors ('correlation' and 'interaction') based on the power-sum function. These regressors were used as nuisance covariates. For a separate post hoc analysis, we defined two PPI regressors ('correlation' and 'interaction') using an average of normalized powers, $[\ln(P(F3))+\ln(P(F7))+\ln(P(FC5))]/3$, for the three EEG channels located over the left prefrontal cortex.

The PPI analysis for each fMRI run involved solution of a GLM model with the PPI regressors by means of the 3dDeconvolve AFNI program. The fMRI data and motion parameters were band-pass filtered between 0.01 and 0.1 Hz. The GLM design matrix included four stimulus regressors, ten nuisance covariates, and five polynomial terms for modeling the baseline. The stimulus regressors included: the FEA-based PPI interaction regressor; the FEA-based PPI correlation regressor; the Happy Memories block-stimulus regressor; the Count block-stimulus regressor. The nuisance covariates included: time courses of the six fMRI motion parameters (together with the same time courses shifted by one *TR*); a time course of a bilateral ROI within ventricle CSF; a time course of a bilateral ROI within white matter; the power-sum-based PPI interaction regressor; the power-sum-based PPI correlation regressor. The last two nuisance covariates accounted for PPI interaction and correlation effects that could be attributed to variations in the average power for the two EEG channels rather than their FEA. We also performed a similar PPI analysis post hoc using the above-mentioned PPI regressors based on the average EEG power over the left prefrontal cortex.

Each PPI analysis yielded GLM-based $R^2$-statistics and *t*-statistics for the PPI interaction and correlation terms, which we used to compute PPI interaction and correlation values for each voxel. The resulting maps were transformed to the Talairach space, re-sampled to 2×2×2 mm$^3$ isotropic voxel size, spatially smoothed (5 mm FWHM), and normalized using Fisher *r*-to-*z* transform. Group *t*-tests with respect to zero level were applied to evaluate significance of the PPI effects. The results were corrected for multiple comparisons by controlling the family-wise-error (FWE). The correction was based on Monte Carlo simulations implemented in the AlphaSim AFNI program.

*2.9. Statistical tests*

Inferential statistical analyses were conducted in IBM SPSS Statistics 20. Statistical significance level for all tests was $\alpha=0.05$, two-tailed. Correction for multiple comparisons was based on controlling the false discovery rate (FDR *q*), which was computed by applying the 3dFDR AFNI program to a column of uncorrected *p*-values from multiple tests. All statistical analyses described below were performed separately for the EG and CG groups, unless stated otherwise.

To test the hypothesis that the MDD patients' relevant emotional states changed after the rtfMRI-nf training, we performed paired *t*-tests on the POMS depression, POMS total mood disturbance, and VAS happiness rating values measured before and after the rtfMRI-nf session. The test results were FDR corrected for multiple comparisons.

To test the hypothesis that the LA was activated during the Happy Memories conditions, we performed *t*-tests (with respect to zero) on Happy Memories vs Rest fMRI percent signal changes for the LA ROI for five task runs (PR, R1, R2, R3, TR). The results were FDR corrected for multiple comparisons over the five runs.

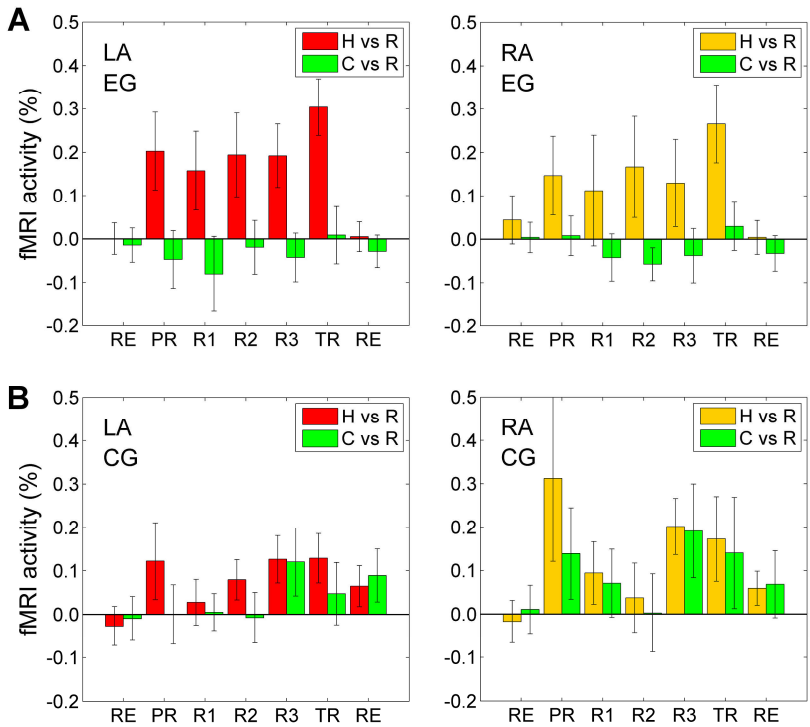

**Figure 3. BOLD activity levels for the amygdala during the rtfMRI-nf experiment. A)** Average fMRI percent signal changes for the left amygdala (LA, *left*) and the right amygdala (RA, *right*) for the experimental group (EG). Each bar represents a mean GLM-based fMRI percent signal change for the corresponding ROI (Fig. 2A) with respect to the Rest baseline for the Happy Memories (H vs R) or Count (C vs R) conditions in a given run, averaged across the group. The error bars are standard errors of the means (sem). The experimental runs and condition blocks are depicted schematically in Fig. 1B. For the Rest runs (RE), the analyses were formally performed in the same way as for the five task runs to evaluate internal consistency of the results. **B)** Corresponding average fMRI percent signal changes for the control group (CG).

To test the hypothesis that the LA BOLD activity levels during the rtfMRI-nf task showed a linear trend across rtfMRI-nf runs (due to the incremental increase in the rtfMRI-nf target level), we performed a one-way ANOVA polynomial trend analysis for repeated measures on Happy Memories vs Rest fMRI percent signal changes for the LA ROI across the runs.

To test the hypothesis that the LA BOLD activity levels during the four rtfMRI-nf runs were different between the EG and CG groups, we applied a two-way 4 (Training) × 2 (Group) between-within mixed factorial repeated measures ANOVA on fMRI percent signal changes for the LA ROI with Training (PR, R1, R2, R3) as a within-subject factor and Group (EG, CG) as a between-subjects factor. Such ANOVAs were conducted separately for the Happy Memories vs Rest activity contrast and for the Happy Memories vs Count activity contrast.

To test the hypothesis that the Happy Memories BOLD activity levels during the five task runs were different between the LA and LHIPS ROIs, we performed a two-way 5 (Training) × 2 (ROI) repeated measures ANOVA on Happy Memories vs Rest fMRI percent signal changes for the two ROIs with Training (PR, R1, R2, R3, TR) and ROI (LA, LHIPS) as within-subjects factors. Such ANOVAs were conducted separately for the EG and CG groups.

To test the hypothesis that average individual BOLD activity levels for a given ROI (LA, RA) during the rtfMRI-nf task correlated with individual psychological measures, we applied Pearson's product-moment correlation analysis to Happy Memories vs Rest fMRI percent signal changes for that ROI (averaged for R1, R2, R3) and individual psychological scores. Similar correlation analyses were performed for the average amygdala BOLD laterality.

The statistical tests described above for the LA BOLD activity levels were conducted to confirm effectiveness of the rtfMRI-nf procedure and compare results to those reported in our previous studies (Zotev et al., 2011; Young et al., 2014). We performed similar statistical tests for average FEA changes between the Rest and Happy Memories conditions to check if similar effects could be observed for FEA.

To test the hypothesis that the Happy Memories vs Rest FEA changes during the four rtfMRI-nf runs were different between the EG and CG groups, we applied a two-way 4 (Training) × 2 (Group) between-within mixed factorial repeated measures ANOVA on the FEA changes with Training (PR, R1, R2, R3) as a within-subject factor and Group (EG, CG) as a between-subjects factor. We also performed a similar two-way 4 × 2 repeated measures ANOVA with the MDD patients' HDRS depression severity included as a covariate.

To test the hypothesis that average individual FEA changes during the rtfMRI-nf task correlated with average individual amygdala BOLD laterality, we applied Pearson's product-moment correlation analysis to Happy Memories vs Rest FEA changes (averaged for PR, R1, R2, R3) and the corresponding amygdala BOLD laterality values (averaged for the same runs). A similar correlation analysis was performed with the data for each of the four runs from each participant included without averaging. The data could be pooled this way, because rtfMRI-nf results typically exhibit large within-subject run-to-run variability comparable to between-subjects variability.

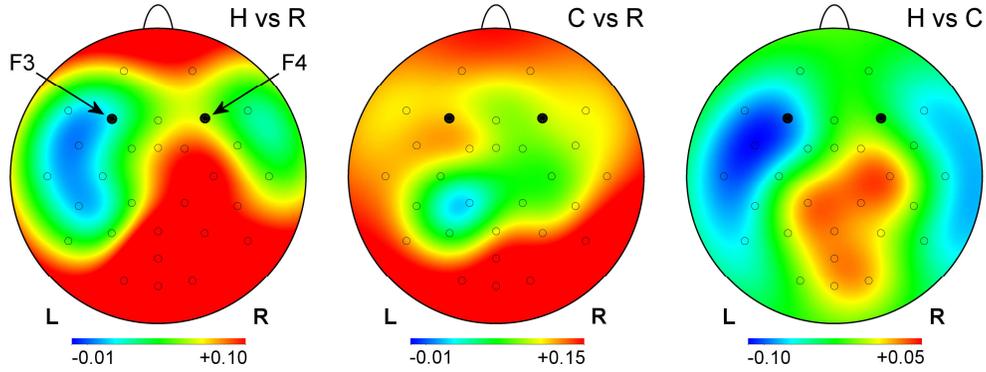

**Figure 4. Variations in upper alpha EEG power across three experimental conditions during the rtfMRI-nf training.** The maps show differences in average normalized upper alpha EEG power values for the Happy Memories and Rest (H vs R), Count and Rest (C vs R), and Happy Memories and Count (H vs C) conditions for the experimental group (EG). The results are also averaged across three rtfMRI-nf training runs (Run 1, Run 2, Run 3). The (unitless) normalized power was computed as $\ln(P)$, where $P$ is EEG signal power for a given channel with Cz reference. The upper alpha EEG band was defined individually for each participant (see text for details).

To test the hypothesis that temporal correlation between FEA and the LA BOLD activity was enhanced during the Happy Memories conditions, we performed $t$-tests (with respect to zero) on the FEA-based PPI interaction values averaged within the LA ROI for five task runs (PR, R1, R2, R3, TR). The results were FDR corrected for multiple comparisons over the five runs.

To test the hypothesis that the FEA-based PPI interaction effect for the LA showed a linear trend across rtfMRI-nf runs, we performed a one-way ANOVA polynomial trend analysis for repeated measures on the PPI interaction (averaged within the LA ROI) across the runs.

To test the hypothesis that the FEA-based PPI interaction values for the LA during the four rtfMRI-nf runs were different between the EG and CG groups, we applied a two-way 4 (Training) × 2 (Group) between-within mixed factorial repeated measures ANOVA on the PPI interaction values (averaged within the LA ROI) with Training (PR, R1, R2, R3) as a within-subject factor and Group (EG, CG) as a between-subjects factor.

## 3. Results

### 3.1. Psychological measures

The MDD patients' emotional state ratings measured before and after the rtfMRI-nf session (with EEG) are reported in Table 1. Three ratings most relevant to the present study – POMS depression, POMS total mood disturbance, and VAS happiness – are included. These emotional state measures showed significant improvements (with FDR correction) after the rtfMRI-nf session for the EG, but not for the CG (Table 1). The changes in the POMS depression ratings showed a significant group difference before correction (EG vs CG: $t(22)=-2.14$, $p<0.045$, $q<0.135$, mean difference $-5.1$, 95% CI $-10.1$ to $-0.13$). The MDD patients' self-report performance ratings (happiness, memory recall) during the rtfMRI-nf session are reported in *Supplementary material* (S2.1).

### 3.2. Amygdala BOLD activity

Figure 3 shows results of the offline fMRI data analysis for the LA and RA ROIs. Note that the average BOLD activity levels for the Happy Memories conditions for the EG tended to be higher for the LA than for the RA (H vs R, Fig. 3A). In contrast, the average activity levels for the CG tended to be higher for the RA than for the LA (H vs R, Fig. 3B). The amygdala activity levels for the CG also tended to be similar for the Happy Memories and Count conditions (Fig. 3B). Results for BOLD activity of the LHIPS region (Fig. 1E) are included in *Supplementary material* (Fig. S2).

Statistics for the LA BOLD activity levels for the EG and CG are reported in detail in *Supplementary material* (S2.2). In particular, the LA BOLD activity levels for the Happy Memories conditions for the EG (H vs R, Fig. 3A) were significant for the Transfer run ($t(12)=4.64$, $q<0.005$). The LA activity levels for the EG also exhibited a positive linear trend ('LT') that was significant across experimental runs (LT(RE...TR): $F(1,12)=9.38$, $p<0.010$).

Correlations between amygdala BOLD activity levels and individual self-report performance ratings and psychological measures are described in detail in *Supplementary material* (S2.3-S2.5). In particular, the average individual LA BOLD activity levels for the Happy Memories conditions (H vs R) for the EG exhibited significant correlations with the average self-report happiness ($r=0.72$, $p<0.006$) and memory-recall ($r=0.73$, $p<0.004$) ratings (S2.3, Fig. S3). The average LA activity levels for the EG also significantly correlated with changes in VAS happiness ($r=0.65$, $p<0.017$) and inversely correlated with changes in POMS tension ($r=-0.62$, $p<0.024$) scores (S2.4, Fig. S4). The corresponding average BOLD activity levels for the RA showed significant inverse correlations with TAS difficulty describing feelings ($r=-0.63$, $p<0.020$) and TAS total score ($r=-0.63$, $p<0.021$) (S2.4, Fig. S4). The average amygdala BOLD laterality ('LA–RA') for the EG exhibited a significant correlation with TAS total scores ($r=0.62$, $p<0.024$) and approaching significance correlation with HDRS depression severity ratings ($r=0.54$, $p<0.057$) (S2.5, Fig. S5).

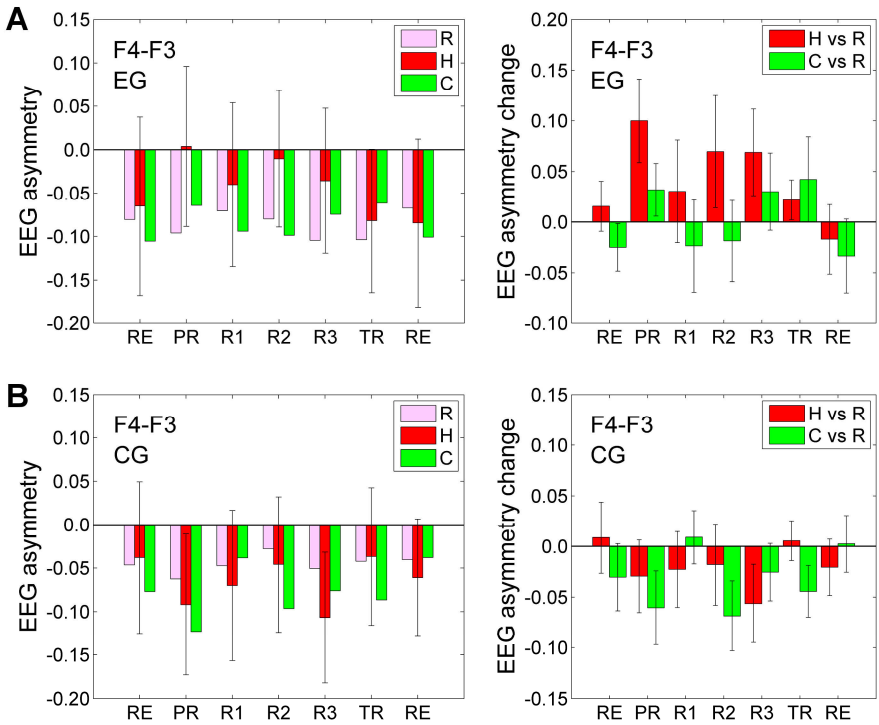

**Figure 5. Average values of frontal EEG asymmetry and their changes during the rtfMRI-nf experiment. A)** Average frontal EEG asymmetry (FEA) values (*left*) and changes (*right*) for channels F3 and F4 (Fig. 2B) for the experimental group (EG). Bars in the left plot represent FEA values averaged, respectively, for the Rest (R), Happy Memories (H), and Count (C) conditions in a given run and across the group. Bars in the right plot represent average FEA changes for the Happy Memories and Count conditions relative to the Rest conditions (H vs R and C vs R). The error bars are standard errors of the means (sem). The notation 'F4–F3' in this and other figures refers to the FEA computed as $\ln(P(F4))-\ln(P(F3))$, where $P$ is EEG signal power for a given channel in an individually defined upper alpha EEG band. The experimental runs and condition blocks are illustrated in Fig. 1B above. For the Rest runs (RE), the analyses were formally performed in the same way as for the five task runs to evaluate internal consistency of the results. **B)** Corresponding average FEA values and changes for the control group (CG).

### 3.3. Frontal EEG asymmetry

Figure 4 shows topographical maps of differences in average normalized upper alpha EEG power between pairs of experimental conditions during the rtfMRI-nf training for the EG. Performance of both the Happy Memories (H) and Count (C) tasks was associated with an overall increase in the upper alpha EEG power compared to the Rest (R) conditions. (Note that the IAF was defined based on the average EEG spectra for the Rest condition blocks). However, the Happy Memories conditions with rtfMRI-nf were characterized by a reduction in the upper alpha EEG power measured by frontal and temporal EEG channels (Fig. 4, H vs R). This reduction was more pronounced for the EEG channels over the left hemisphere (F3, F7, FC5, T7) than for their counterparts on the right. For the Count conditions, the average upper alpha EEG power was reduced for parietal EEG channels (Fig. 4, C vs R), also with left lateralization (CP1, P3). These effects were also evident for the Happy Memories vs Count condition contrast (Fig. 4, H vs C).

Average FEA values for channels F3 and F4 (denoted 'F4–F3') and their changes across the three experimental conditions are shown in Fig. 5. The average FEA changes for the Happy Memories conditions with rtfMRI-nf compared to the Rest conditions were positive for the EG (H vs R, Fig. 5A, right) and negative for the CG (H vs R, Fig. 5B, right). The results for individual runs were not significant after an FDR correction for reasons explained below. A two-way 4 (Training: PR, R1, R2, R3) × 2 (Group: EG, CG) repeated measures ANOVA applied to the Happy Memories vs Rest FEA changes (H vs R, Fig. 5, right) revealed an effect for the Group that showed a trend toward significance ($F(1,22)=3.74$, $p<0.066$). A similar two-way 4 × 2 repeated measures ANOVA with the HDRS depression severity included as a covariate showed a significant effect of the Group ($F(1,21)=5.80$, $p<0.025$) and a significant effect of the HDRS covariate ($F(1,21)=8.78$, $p<0.007$).

### 3.4. EEG asymmetry changes vs depression severity

Figure 6 demonstrates correlations between the average FEA changes during the rtfMRI-nf training and individual psychological trait measures for the EG participants. Remarkably, the average FEA changes during the Happy Memories conditions with rtfMRI-nf relative to the Rest conditions (H vs R) showed significant positive correlations with both depression severity (HDRS) and anhedonia severity (SHAPS) ratings. These ratings also correlated with each other (SHAPS vs HDRS:

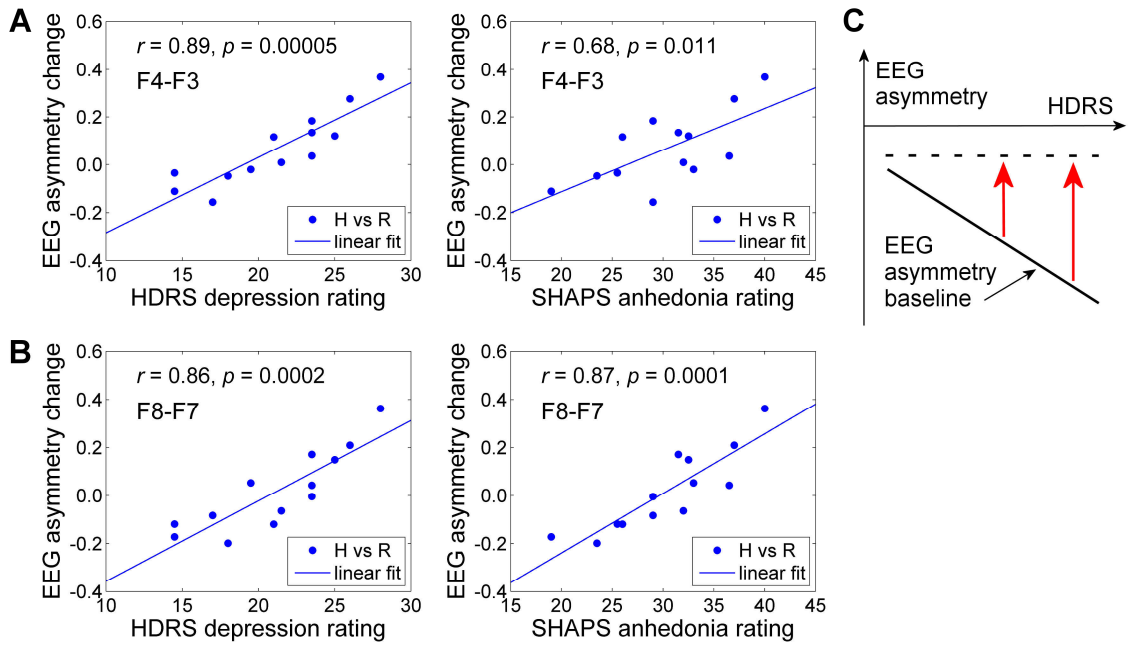

**Figure 6. Correlations between frontal EEG asymmetry changes during the rtfMRI-nf training and individual psychological measures.** The results are for the experimental group (EG), with each data point corresponding to one participant. Mean frontal EEG asymmetry (FEA) changes for the Happy Memories conditions with respect to the Rest conditions (H vs R) further averaged across four rtfMRI-nf runs (Practice, Run 1, Run 2, Run 3) were used in the analysis for each subject. **A)** Correlation results for the FEA changes for channels F3 and F4 ('F4–F3'). **B)** Correlation results for the FEA changes for channels F7 and F8 ('F8–F7'). **C)** Interpretation of the experimental results. The baseline FEA values (Rest condition blocks) are more negative (solid line) in patients with higher depression severity (HDRS). The rtfMRI-nf FEA values (Happy Memories conditions) appear independent (dashed line) of the depression severity. Thus, the FEA changes (red arrows) are more positive in patients with more severe depression. Abbreviations: HDRS – Hamilton Depression Rating Scale, SHAPS – Snaith-Hamilton Pleasure Scale.

$r=0.84$, $p<0.0004$). It should be noted that the *positive* correlations between the FEA *changes* and depression severity ratings (HDRS) in Figs. 6A,B primarily reflect the *inverse* correlations between the baseline FEA *values* (for the Rest condition blocks) and the depression severity. Indeed, the average Rest FEA values, corresponding to the results in Figs. 6A,B, inversely correlated with the HDRS ratings ($r=-0.34$, $p<0.259$ for F4 vs F3; $r=-0.52$, $p<0.073$ for F8 vs F7). Correlations between the average Happy Memories FEA values and the same ratings were considerably weaker ($r=0.07$, $p<0.826$ for F4 vs F3; $r=-0.13$, $p<0.673$ for F8 vs F7). This interpretation is illustrated schematically in Fig. 6C.

Figure 7 reveals a connection between an enhancement in EEG coherence across rtfMRI-nf runs and individual depression severity. We defined the EEG coherence slope (ECS) as a slope of the linear fit to the average upper alpha EEG coherence values for the Happy Memories conditions across four rtfMRI-nf runs (Fig. 7B). For the EG participants, the ECS values for the left fronto-temporal EEG channel pairs exhibited positive correlations ($p<0.05$, uncorr.) with the HDRS depression severity ratings (Figs. 7A,C). Notably, the average individual ECS laterality, ECS(L)–ECS(R), i.e. the difference between mean ECS values for the fronto-temporal EEG channel pairs on the left and on the right (Fig. 7D), showed a strong positive correlation with the HDRS ratings ($r=0.83$, $p<0.0004$). The corresponding lateral ECS vs HDRS correlations were less significant and had opposite signs (ECS(L) vs HDRS: $r=0.67$, $p<0.012$; ECS(R) vs HDRS: $r=-0.21$, $p<0.487$).

### 3.5. Amygdala-asymmetry correlations

The average FEA changes during the rtfMRI-nf training (H vs R) for the EG exhibited significant positive correlations with the average amygdala BOLD laterality, as demonstrated in Fig. 8. The correlations were also significant when the EG data were pooled across the subjects and runs ($r=0.47$, $n=52$, $p<0.0005$ for F4 vs F3; $r=0.51$, $n=52$, $p<0.0001$ for F8 vs F7). No significant correlations were observed between the same FEA changes and BOLD activity levels for either the LA or RA separately. For the CG, correlations between the average FEA changes and the amygdala BOLD laterality corresponding to the results in Fig. 8 were not significant (CG: $r=-0.32$, $p<0.339$ for F4 vs F3; $r=-0.11$, $p<0.753$ for F8 vs F7). Similarly, no correlations were found when the CG data were pooled across the subjects and runs (CG: $r=-0.02$, $n=44$, $p<0.916$ for F4 vs F3; $r=-0.03$, $n=44$, $p<0.833$ for F8 vs F7).

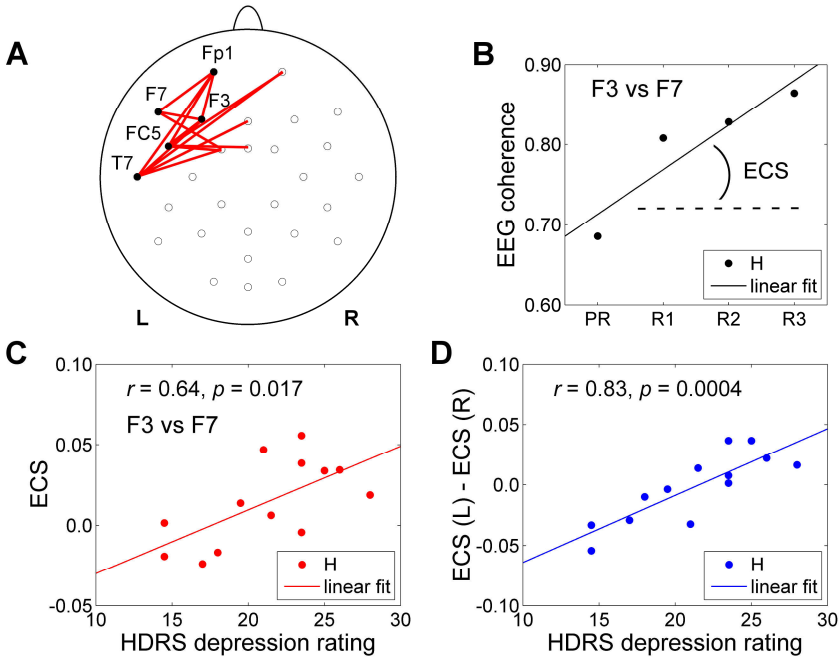

**Figure 7. Correlations between enhancement in EEG coherence in the upper alpha EEG band during the rtfMRI-nf training and depression severity. A)** Pairs of EEG channels that showed correlations ($p<0.05$, uncorr.) between the EEG coherence slope (ECS) and HDRS depression severity for the experimental group (EG). **B)** Definition of the EEG coherence slope (ECS) across Happy Memories conditions in four rtfMRI-nf runs (Practice, Run 1, Run 2, Run 3) for a given pair of EEG channels for a given subject. **C)** Example of correlation between the ECS values (for F3 vs F7) and HDRS depression severity ratings for the participants in the EG. Each data point corresponds to one participant. Pairs of EEG channels exhibiting such correlations are denoted by red segments in A). **D)** Correlation between the average ECS laterality and HDRS depression severity ratings. ECS(L) is a mean ECS for fronto-temporal EEG channels on the left (Fp1, F3, F7, FC5, T7, ten pairs). ECS(R) is a mean ECS for the corresponding fronto-temporal EEG channels on the right (Fp2, F4, F8, FC6, T8, ten pairs). HDRS – Hamilton Depression Rating Scale.

Figure 9 shows results of the PPI analyses based on the FEA for channels F3 and F4, averaged within the LA ROI for the EG (Fig. 9A) and for the CG (Fig. 9B). Average values of the PPI interaction for the EG were positive for Runs 1-3 and the Transfer run (Fig. 9A). The PPI interaction effect was significant for Run 3 ($t(12)=3.03$, $q<0.050$). It also exhibited a significant positive linear trend across experimental runs (LT(RE...R3): $F(1,12)=10.18$, $p<0.008$; LT(RE...TR): $F(1,12)=7.39$, $p<0.019$). There was no significant difference in the mean PPI interaction values between Run 3 and the Transfer run for the EG (TR vs R3: $t(12)=-0.76$, $p<0.463$). Average values of the PPI interaction for the CG were negative for the four rtfMRI-nf runs (Fig. 9B). A two-way 4 (Training: PR, R1, R2, R3) × 2 (Group: EG, CG) repeated measures ANOVA applied to the PPI interaction values (Figs. 9A,B) revealed a significant effect for the Group ($F(1,22)=6.36$, $p<0.019$) and a Training×Group interaction effect that showed a trend toward significance ($F(3,66)=2.70$, $p<0.052$). PPI effects for the EG are illustrated in Fig. 9C using single-subject data for a single run.

Whole-brain group statistical maps of the PPI interaction effect for the EG are exhibited in Fig. 10. The PPI results for only one rtfMRI-nf training run (among Runs 1-3), characterized by the largest positive average FEA change between the Rest and Happy Memories conditions (H vs R), were included in the group analysis from each participant. The group statistical maps for the PPI interaction in Fig. 10 were thresholded at $t=\pm3.06$ (uncorr. $p<0.01$) and clusters containing at least 75 voxels (corr. $p<0.05$) are shown in the figure. For a more accurate localization of the PPI effects, the same data were thresholded at $t=\pm4.32$ (uncorr. $p<0.001$) and clusters containing at least 24 voxels (corr. $p<0.05$) are described in Table 2. The results in Fig. 10 and Table 2 demonstrate that the left amygdala and various other brain regions involved in emotion regulation exhibited significant positive PPI interaction effects. This means that the temporal correlation between their BOLD activity and the FEA for channels F3 and F4 (convolved with the HRF) was significantly enhanced during the Happy Memories conditions with rtfMRI-nf compared to the Count conditions.

Figure 11 exhibits results of a post hoc PPI analysis conducted using an average upper alpha EEG power over the left prefrontal cortex. Because task-dependent variations in upper alpha EEG power were similar for channels F3, F7, and FC5 for the EG (Fig. 4), we defined PPI regressors using an average of their normalized powers. Group statistical maps for the PPI interaction effect were thresholded at $t=\pm3.06$ (uncorr. $p<0.01$) and clusters containing at least 75 voxels (corr. $p<0.05$) were retained. Three clusters emerged in the analysis showing significant negative PPI interactions (Fig. 11, Table 3). The strongest effect was observed for the left DM/DLPFC cluster with the $t$-score peak at ($-11$, 29, 40) in the left medial frontal gyrus (BA 8) and the center of mass at ($-13$, 29, 43) in the superior frontal gyrus (BA 8). The results in Fig. 11 and Table 3 indicate a significant enhancement in *inverse* temporal correlation between the upper alpha EEG power over the left prefrontal cortex (convolved with the HRF) and BOLD activity in those three regions during the rtfMRI-nf task.

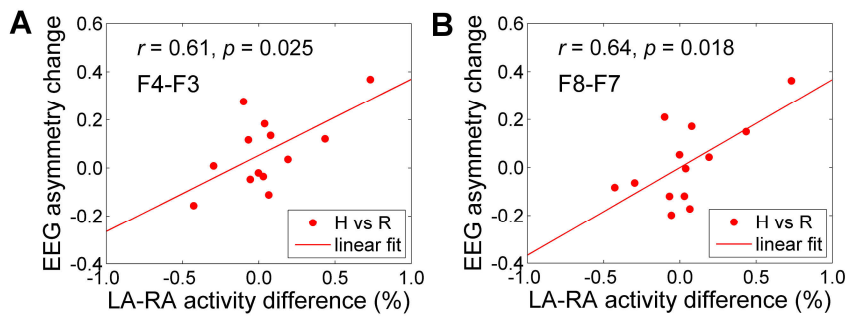

**Figure 8. Correlations between changes in frontal EEG asymmetry and amygdala BOLD laterality during the rtfMRI-nf training.** The results are for the experimental group (EG), with each data point corresponding to one participant. Mean individual frontal EEG asymmetry (FEA) changes and BOLD activity differences between the LA and RA ('LA–RA') for the Happy Memories conditions with respect to the Rest conditions (H vs R) were averaged across four rtfMRI-nf runs (Practice, Run 1, Run 2, Run 3). **A)** FEA changes for channels F3 and F4 (denoted as 'F4–F3'). **B)** FEA changes for channels F7 and F8 (denoted as 'F8–F7').

## 4. Discussion

The MDD patients in the present study were able to upregulate their amygdala BOLD activity by means of the rtfMRI-nf during happy emotion induction. We aimed to evaluate electrophysiological effects of such rtfMRI-nf training by conducting simultaneous (passive) EEG recordings. With this approach, we specifically examined correlations between the FEA in the upper alpha band and BOLD activity across the brain during the rtfMRI-nf procedure. Our approach is novel and revealing, because it enables an independent EEG-based evaluation of the rtfMRI-nf effects.

Following the rtfMRI-nf session, the MDD patients in the EG showed significant reductions in state depression and total mood disturbance, as well as a significant increase in state happiness (Table 1). These results are consistent with those reported previously for the first rtfMRI-nf session in Young et al., 2014. The readers are referred to that work for a discussion of psychological measures. Our experimental results concerning specifically the rtfMRI-nf performance and the amygdala BOLD activity are discussed in detail in *Supplementary material* (S3.1-S3.4).

To evaluate electrophysiological correlates of the rtfMRI-nf training we examined changes in the FEA across the experimental conditions. The average FEA changes for the Happy Memories conditions with rtfMRI-nf (relative to the Rest conditions) tended to be positive for the EG (Fig. 5A) and negative for the CG (Fig. 5B). These FEA effects are consistent with predictions of the approach-withdrawal hypothesis. Happiness is associated with approach motivation and more positive FEA values (Coan et al., 2001; Davidson et al., 1990). This approach motivation might conceivably have been further enhanced during the performance of the rtfMRI-nf task for the EG, because successful upregulation of the rtfMRI-nf signal requires an active emotional engagement and achievement motivation on the part of the participant. Indeed, the average FEA changes for the EG (Fig. 5A, right) were larger for the four rtfMRI-nf runs than for the Transfer run, suggesting that these changes were associated in part with the rtfMRI-nf and could not be attributed solely to the happy emotion induction. For the CG, however, the sham rtfMRI-nf provided information inconsistent with performance of the emotion induction task. The participant's inability to control the rtfMRI-nf signal in this case might have diminished the approach motivation. It might also have led to increased anxiety and stress, which are associated with avoidance motivation and more negative FEA values (Davidson et al., 2000; Lewis et al., 2007). The average FEA changes for the EG (Fig. 5A, right) were not significant after an FDR correction. The primary reason for this appears to be the distribution of depressive symptoms which led to the strong dispersion in individual FEA changes (Fig. 6). Furthermore, self-regulation of the amygdala activity using rtfMRI-nf is a difficult task, so the feelings of stress and anxiety could have been experienced, to some extent, by the EG participants as well. Importantly, the group difference in the FEA changes between the EG and CG showed a trend toward significance ($p<0.066$), and was significant ($p<0.025$) when the HDRS variability was explicitly taken into account (Sec. 3.3). These results support the first of our main hypotheses (Sec. 1) indicating the neurofeedback-specific enhancement in approach motivation.

An important result of our study is the observation of the significant *positive* correlations between the average individual FEA changes for the Happy Memories conditions with rtfMRI-nf (relative to the Rest conditions) and the EG participants' depression (HDRS) and anhedonia (SHAPS) ratings (Figs. 6A,B). This observation suggests that the rtfMRI-nf training may have stronger therapeutic effects, in terms of increasing approach motivation, in MDD patients with more severe depression. We emphasize that this result does not mean that the MDD patients with higher depression severity achieved higher FEA levels during the rtfMRI-nf task. In fact, the average FEA values for the Happy Memories conditions with rtfMRI-nf did not exhibit any correlations with the individual HDRS scores. In contrast, the average FEA values for the Rest conditions showed *inverse* correlations with the HDRS ratings. Therefore, the MDD patients with higher depression severity exhibited more positive FEA *changes* mainly because their baseline FEA *values* were more negative (Fig. 6C). The last observation is consistent with the results of numerous EEG studies that demonstrated an inverse relationship between severity of MDD symptoms and FEA values at rest.

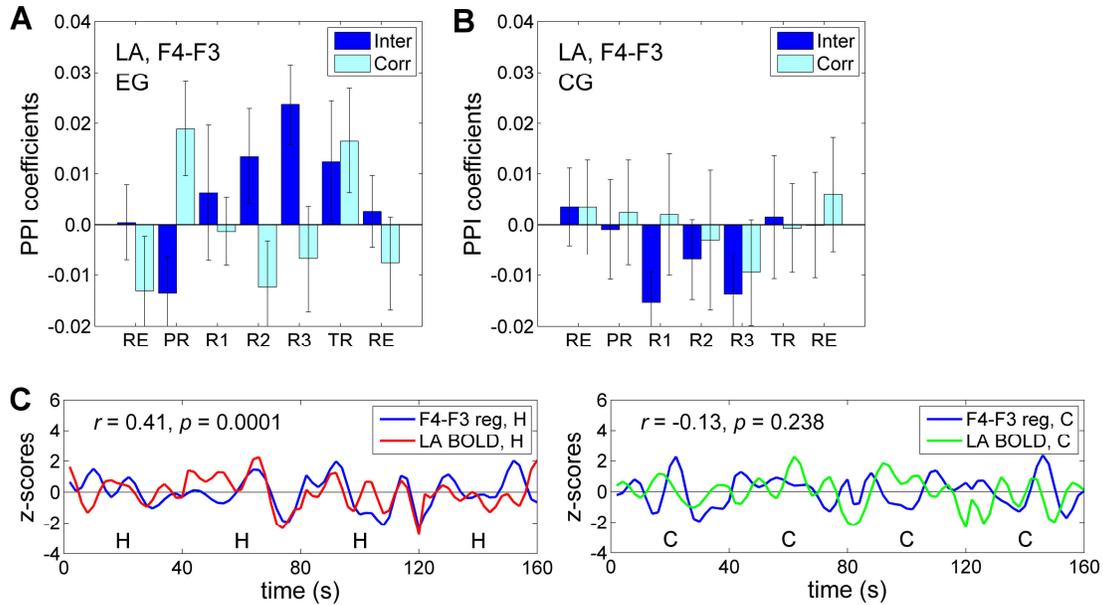

Figure 9. Temporal correlation between frontal EEG asymmetry and BOLD activity of the left amygdala during the rtfMRI-nf training. **A)** Average values of the psychophysiological interaction (PPI) analysis coefficients – interaction ('Inter') and correlation ('Corr') – for the left amygdala (LA) ROI (Fig. 2A) for the experimental group (EG). Frontal EEG asymmetry (FEA) for channels F3 and F4 (denoted as 'F4–F3') was used in the PPI analysis. The voxel-wise [FEA-based regressor] × [Happy−Count] interaction and correlation values were averaged within the LA ROI and across the group. The error bars are standard errors of the means (sem). The positive PPI interaction indicates a stronger temporal correlation between the FEA and the LA BOLD activity during the Happy Memories condition. **B)** Corresponding PPI results for the control group (CG). **C)** Illustration of the PPI effects for the EG using single-subject data. The left plot shows positive correlation between the FEA-based regressor and the LA time course during four Happy Memories (H) condition blocks in one run (Fig. 1B) concatenated together in the figure. The right plot shows lack of correlation between these time courses during four concatenated Count (C) condition blocks in the same run.

Another interesting finding is that the MDD patients with more severe depression showed *stronger* enhancement in the upper alpha EEG coherence for the left fronto-temporal EEG channels across the rtfMRI-nf runs (Fig. 7). Such EEG coherence enhancement indicates an enhancement in functional connectivity among the underlying left fronto-temporal cortical regions. The laterality of the enhancement in EEG coherence showed a significant positive correlation with the HDRS ratings (Fig. 7D). Remarkably, we observed a similar effect in our fMRI-only analysis of the same multimodal data (Zotev et al., 2015). The laterality of the enhancement in fMRI functional connectivity between the left/right DLPFC (middle frontal gyrus, BA 8) and the left amygdala during the rtfMRI-nf training also exhibited a significant positive correlation with the HDRS ratings (Zotev et al., 2015, Fig. 2D therein). Thus, the MDD-specific laterality of functional connectivity changes during the rtfMRI-nf training can be observed independently in both EEG and fMRI data.

A novel contribution of the present work is the multimodal investigation of correlations between the amygdala BOLD activity and the FEA. The average individual FEA changes during the rtfMRI-nf task showed significant positive correlations with the average individual amygdala BOLD laterality ('LA–RA') for the EG (Fig. 8), but not for the CG. The results in Fig. 8 can be interpreted as follows. The amygdala BOLD laterality in the present study can be viewed as a measure of *target-specific* rtfMRI-nf effects, as discussed in *Supplementary material* (S1.2). The FEA changes also reflect performance of the rtfMRI-nf task, as argued above. Therefore, Fig. 8 shows positive correlations between the *neurofeedback-specific* variations in the amygdala BOLD activity and the *neurofeedback-specific* changes in the FEA. This result suggests that a stronger approach motivation during the rtfMRI-nf task was associated with a higher relative activity of the target amygdala region (LA vs RA), i.e. a larger amygdala BOLD laterality.

The results of the PPI analyses (Fig. 9) demonstrate that performance of the rtfMRI-nf task was associated with a significant enhancement in temporal correlation between the FEA and the LA BOLD activity for the EG (Fig. 9A), but not for the CG (Fig. 9B). Moreover, such enhancement showed a significant positive linear trend across the experimental runs and generalized beyond the actual rtfMRI-nf training (Fig. 9A). These results support the second of our main hypotheses (Sec. 1).

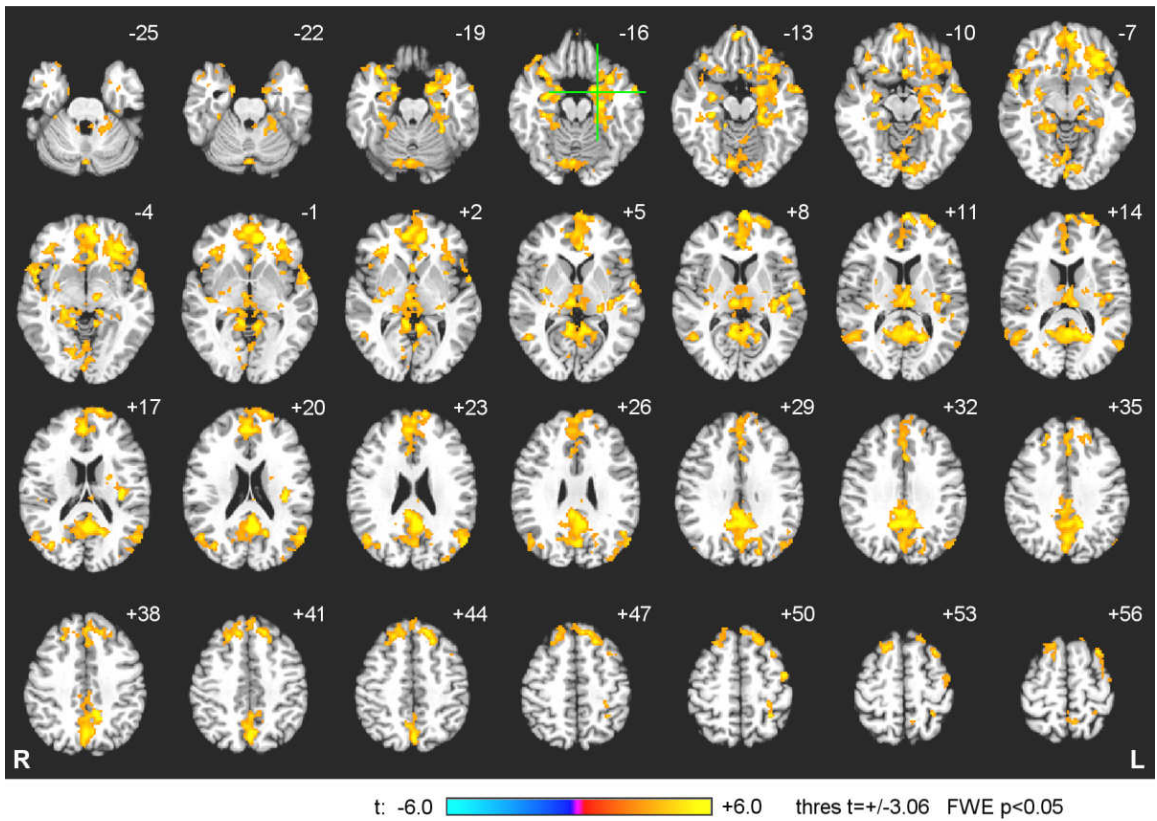

**Figure 10. Enhancement in temporal correlation between frontal EEG asymmetry and BOLD activity of various brain regions during the rtfMRI-nf training.** Group statistical maps of the psychophysiological interaction (PPI) effect [FEA-based regressor] × [Happy−Count] are shown for the experimental group (EG). Frontal EEG asymmetry (FEA) for channels F3 and F4 was used in the PPI analysis. The maps are projected onto the standard anatomical template TT_N27 in the Talairach space, with 3 mm separation between axial slices. The number adjacent to each slice indicates the $z$ coordinate in mm. The left hemisphere (L) is to the reader's right. The green crosshairs mark the center of the left amygdala target ROI (Fig. 1D). Peak $t$-statistics values for the PPI interaction effect and the corresponding locations are specified in Table 2.

The whole-brain group PPI results (Fig. 10, Table 2) further demonstrate that the temporal correlation between the FEA and BOLD activity during the rtfMRI-nf task was significantly enhanced not only for the LA, but also for a broad network of regions commonly involved in emotion regulation. The most significant PPI interaction effect in the left amygdala region was observed at the (−17, −3, −16) locus (Table 2). This location is spatially close to the (−17, −7, −16) point, which exhibited the largest BOLD activity contrast between the EG and CG in our rtfMRI-nf study on healthy participants (Zotev et al., 2011). That point was used as a center of the LA seed ROI in our functional and effective connectivity analyses (Zotev et al., 2011, 2013a). Comparison of the PPI results in Fig. 10 and Table 2 with the LA functional connectivity results in Zotev et al., 2011 (Fig. 8 and Table 3 therein) is revealing. The peak $t$-score locations in the two studies are spatially close (<12 mm apart) for the left lateral orbitofrontal cortex (OFC, BA 47), the right medial frontopolar cortex (BA 9/10), the left middle temporal gyrus (BA 39 and BA 21), the left rostral (pregenual) anterior cingulate cortex (ACC, BA 24), the left amygdala, the right amygdala, the right hippocampus, the right parahippocampal gyrus (BA 36), and the left insula (BA 13). Such close spatial correspondence of the results from two different modalities (EEG vs fMRI), two different analyses (FEA-based PPI vs fMRI functional connectivity), and two different studies (MDD patients vs healthy subjects) suggests that the FEA is a meaningful temporal measure of emotion/motivation that may indirectly reflect activity of the amygdala.

The whole-brain group PPI results support the hypothesis behind the PPI analysis in the present work (Sec. 2.8). Significant positive PPI interaction effects (Fig. 10, Table 2) are observed for several brain regions that have been consistently associated with approach and/or avoidance motivation in fMRI-only studies (e.g. Spielberg et al., 2011, 2012). These regions include the left and right DLPFC (BA 8, middle frontal gyrus, superior frontal gyrus), the left and right lateral OFC (BA 47, inferior frontal gyrus), the left rostral ACC (BA 32/24), the left and right amygdala (Table 2). The prominent involvement of these regions suggests that it is modulation of the motivation networks involving the amygdala during the rtfMRI-nf procedure that leads to the observed correlations between the FEA and the amygdala BOLD activity.

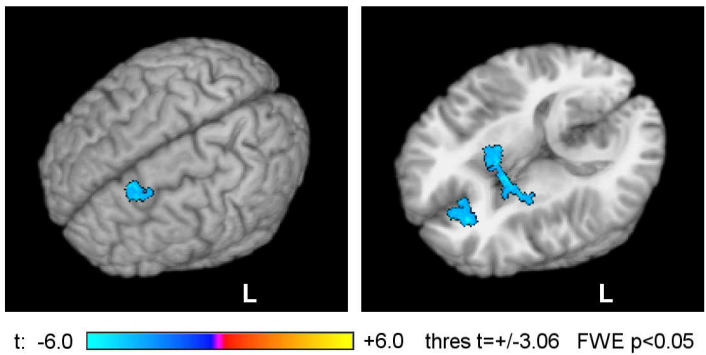

**Figure 11. Enhancement in inverse temporal correlation between left frontal upper alpha EEG power and BOLD activity during the rtfMRI-nf training.** 3D renderings of group statistical maps of the psychophysiological interaction (PPI) effect [EEG-power-based regressor] × [Happy–Count] are shown for the experimental group (EG). An average of normalized upper alpha EEG powers for channels F3, F7, and FC5 located over the left prefrontal cortex was used in the PPI analysis. The maps are projected onto the standard 3D anatomical template TT_N27 in the Talairach space. Three clusters, obtained after FWE correction, are shown as follows. *Left*: a cluster in the left DM/DLPFC region. *Right*: a cluster in the left rACC region and a cluster including areas of the right globus pallidus and the bilateral amygdala (show-through 'glass brain' rendering). Peak *t*-statistics values for the PPI interaction effect and the corresponding locations are specified in Table 3.

While the FEA-based PPI results in Fig. 10 and Table 2 reflect differences between approach and avoidance motivation, the PPI results in Fig. 11 and Table 3 should more specifically reflect variations in approach motivation. These results indicate that performance of the rtfMRI-nf task was associated with a significant enhancement in inverse temporal correlation between the average upper alpha EEG activity over the left prefrontal cortex and BOLD activity in three distinct regions. These regions – the left DLPFC, the left rACC, and the bilateral amygdala together with the globus pallidus – are indeed parts of the approach motivation system, as suggested by fMRI-only studies (Spielberg et al., 2011, 2012). The results in Fig. 11 show that by monitoring upper alpha EEG activity over the left prefrontal cortex during the rtfMRI-nf procedure one can indirectly probe activities of the left rACC and the bilateral amygdala. Note that our effective connectivity analysis of the rtfMRI-nf data for healthy participants suggested that it was the left rACC that modulated activities of the LA and several prefrontal cortical regions during the rtfMRI-nf training (Zotev et al., 2013a). Furthermore, the left rACC showed an enhancement in fMRI functional connectivity with the LA during the rtfMRI-nf training that significantly correlated with the MDD patients' HDRS ratings (Zotev et al., 2015).

A limitation of the present study results from the complexity of the main experimental task, which combines happy emotion induction, retrieval of autobiographical memories, and regulation of the rtfMRI-nf signal – all performed simultaneously in real time. Because of this complexity, effects of emotion, motivation, episodic memory, and executive function on the FEA cannot be reliably separated and evaluated independently. This limitation, however, is not prohibitive for the following reasons. *First*, episodic memory retrieval is associated with greater activity of the right prefrontal regions (e.g. Kalpouzos & Nyberg, 2010). Sustained vigilance is also lateralized to the right hemisphere (e.g. Heilman, 1996). These processes during the rtfMRI-nf task can be expected to lower the FEA, so they cannot explain the positive FEA changes for the EG (Fig. 5A). *Second*, the average FEA changes during the rtfMRI-nf task showed a significant correlation with the EG participants' HDRS depression severity ratings (Fig. 6). This means that these FEA changes, irrespective of relative contributions of different underlying processes, depend on and provide important information about the severity of MDD symptoms. *Third*, the FEA as a function of time exhibited a significantly enhanced correlation with BOLD activity of the amygdala and many regions of the emotion regulation network during the rtfMRI-nf task for the EG (Fig. 10). This finding indicates that temporal FEA variations in our experiments are strongly related to emotion regulation.

The rtfMRI-nf with simultaneous EEG approach, described in this paper, complements the EEG-nf with simultaneous fMRI approach, which has been used by several groups (e.g. Cavazza et al., 2014; Kinreich et al., 2014; Meir-Hasson et al., 2014; Shtark et al., 2015; Zich et al., 2015). The latter method enables evaluation and validation of various EEG-nf techniques using simultaneously acquired fMRI data. A major technical challenge facing the EEG-nf with simultaneous fMRI is the need for an accurate removal of EEG-fMRI artifacts in real time to ensure that the EEG-nf signal provided to a subject inside an MRI scanner reflect the actual neuronal activity, not artifacts. The rtfMRI-nf with simultaneous EEG method, used in the present study, allows direct rtfMRI-nf modulation of small precisely defined regions deep inside the brain, such as the amygdala or individual thalamic nuclei, and investigation of related EEG activity. This method can help to better understand connections between BOLD activities of particular brain regions and networks and various EEG activity patterns and metrics. The problem of EEG-fMRI artifacts does not play a significant role in this approach, because the entire EEG data analysis is performed offline, and the artifacts can be accurately removed using the most advanced techniques available.

One recent neuromodulation study (Cavazza et al., 2014; Gilroy et al., 2013) employed an EEG-nf based on the FEA in the alpha band with simultaneous fMRI. The authors reported an increase in BOLD fMRI activity in the prefrontal cortex during an emotional support task with upregulation of the FEA using the EEG-nf (Fig. 5 in Cavazza et al., 2014). This result provided an initial fMRI validation of the FEA-based EEG-nf. However, the study did not include a control group (needed to prove specificity of the observed BOLD activity to the EEG-nf), and did not involve any EEG-fMRI correlation analysis. No temporal correlations between the FEA time courses and BOLD activity time courses were examined, and no correlations between mean FEA changes and mean BOLD activity levels were evaluated. No results for the amygdala region were reported. Therefore, that study, though interesting and encouraging, did not directly relate the FEA in the alpha band to BOLD activity. Such relation was initially reported in Zotev et al., 2013b, 2013c, and is investigated in detail in the present paper (Figs. 8, 9, 10).

Overall, our results suggest that the upregulation of the amygdala BOLD activity using the rtfMRI-nf during the happy emotion induction task is beneficial to MDD patients. We emphasize that this conclusion is based on the observed FEA and EEG coherence variations, rather than psychological changes as measured by clinical scales or behavioral testing. It is the interpretation of the FEA within the approach-avoidance framework (Sec. 1) that makes it possible to relate the FEA variations to potential clinical and behavioral benefits. Deficient approach motivation, associated with diminished abilities to experience positive affect and engage in goal-oriented behaviors, is a major neuropsychological impairment in depression. Our results show that the average FEA changes during the rtfMRI-nf task were more positive for the EG than for the CG (Fig. 5), suggesting enhancement in approach motivation specific to the amygdala rtfMRI-nf. Remarkably, the FEA changes for the EG were more positive in the MDD patients with more severe depression (Fig. 6). Similarly, the enhancement in upper alpha EEG coherence for the EG showed a greater left-right laterality in the MDD patients with higher depression severity (Fig. 7D). These results suggest that the described rtfMRI-nf procedure may be effective in both moderately and severely depressed MDD patients, and may have the ability to correct the functional impairments specific to major depressive disorder.

## 5. Conclusion

We demonstrated that EEG recordings performed simultaneously with the rtfMRI-nf training of emotion regulation can provide important real-time information about the participants' emotional states. Our EEG data analysis suggests that the rtfMRI-nf training targeting the amygdala is beneficial to MDD patients. We observed, for the first time, positive temporal correlation between the FEA in the alpha band and the amygdala BOLD activity. This finding retrospectively validates, supports, and justifies the use of the FEA as a measure of emotion/motivation in numerous previous EEG studies. It also suggests that EEG-nf aimed at increasing the FEA in the alpha band would be compatible with the amygdala-based rtfMRI-nf. The two types of neurofeedback would complement each other, and could be used either separately during alternating EEG-nf and rtfMRI-nf sessions, or simultaneously as the rtfMRI-EEG-nf. Such multimodal neuromodulation could enhance efficiency of emotion regulation training in patients with depression.

## Acknowledgments

This work was supported by the Laureate Institute for Brain Research and the William K. Warren Foundation. We would like to thank Dr. Robert Störmer, Dr. Patrick Britz, Dr. Markus Plank, and Dr. Mario Bartolo of Brain Products, GmbH for their help and technical support.

## References

Aftanas, L.I., Golocheikine, S.A., 2001. Human anterior and frontal midline theta and lower alpha reflect emotionally positive state and internalized attention: high-resolution EEG investigation of meditation. Neurosci. Lett. 310, 57-60.

Allen, J.J.B., Harmon-Jones, E., Cavender, J.H., 2001. Manipulation of frontal EEG asymmetry through biofeedback alters self-reported emotional responses and facial EMG. Psychophysiol. 38, 685-693.

American Psychiatric Association, 2000. Diagnostic and Statistical Manual of Mental Disorders, 4th ed. text rev. (DSM-IV-TR). American Psychiatric Press, Washington, DC.

Bagby, R.M., Parker, J.D.A., Taylor, G.J., 1994. The twenty-item Toronto Alexithymia Scale – I. Item selection and cross-validation of the factor structure. J. Psychosom. Res. 38, 33-40.

Baehr, E., Rosenfeld, J.P., Baehr, R., 1997. The clinical use of an alpha asymmetry protocol in the neurofeedback treatment of depression: two case studies. J. Neurotherapy, 2, 10-23.

Bell, A.J., Sejnowski, T.J., 1995. An information-maximization approach to blind separation and blind deconvolution. Neural Comput. 7, 1129-1159.

Berkman, E.T., Lieberman, M.D., 2010. Approaching the bad and avoiding the good: lateral prefrontal cortical asymmetry distinguishes between action and valence. J. Cogn. Neurosci. 22, 1970-1979.

Bodurka, J., Bandettini, P.A., 2008. Real time software for monitoring MRI scanner operation. Proc. Org. Hum. Brain Mapping (OHBM), NeuroImage 41 (Suppl. 1), S85.

Canli, T., Desmond, J.E., Zhao, Z., Glover, G., Gabrieli, J.D.E., 1998. Hemispheric asymmetry for emotional stimuli detected with fMRI. NeuroReport 9, 3233-3239.

Caria, A., Veit, R., Sitaram, R., Lotze, M., Weiskopf, N., Grodd, W., Birbaumer, N., 2007. Regulation of anterior insular cortex activity using real-time fMRI. NeuroImage 35, 1238-1246.

Caria, A., Sitaram, R., Veit, R., Begliomini, C., Birbaumer, N., 2010. Volitional control of anterior insula activity modulates the response to aversive stimuli. A real-time functional magnetic resonance imaging study. Biol. Psychiatry 68, 425-432.

Cavazza, M., Aranyi, G., Charles, F., Porteous, J., Gilroy, S., Klovatch, I., Jackont, G., Soreq, E., Keynan, N.J., Cohen, A., Raz, G., Hendler, T., 2014. Towards empathic neurofeedback for interactive storytelling. OpenAccess Ser. Informatics 41, 42-60.

Choi, S.W., Chi, S.E., Chung, S.Y., Kim, J.W., Ahn, C.Y., Kim, H.T., 2011. Is alpha wave neurofeedback effective with randomized clinical trials in depression? A pilot study. Neuropsychobiol. 63, 43-51.

Coan, J.A., Allen, J.J.B., 2004. The state and trait nature of frontal EEG asymmetry in emotion. In: Hugdahl, K., Davidson, R.J. (Eds.), The Asymmetrical Brain. The MIT Press, Cambridge, MA. 565-615.

Coan, J.A., Allen, J.J.B., Harmon-Jones, E., 2001. Voluntary facial expression and hemispheric asymmetry over the frontal cortex. Psychophysiol. 38, 912-925.

Cox, R.W., 1996. AFNI: software for analysis and visualization of functional magnetic resonance neuroimages. Comput. Biomed. Res. 29, 162-173.

Cox, R.W., Hyde, J.S., 1997. Software tools for analysis and visualization of fMRI data. NMR Biomed. 10, 171-178.

Cunningham, W.A., Raye, C.L., Johnson, M.K., 2005. Neural correlates of evaluation associated with promotion and prevention regulatory focus. Cogn. Affect. Behav. Neurosci. 5, 202-211.

Cunningham, W.A., Arbuckle, N.L., Jahn, A., Mowrer, S.M., Abduljalil, A.M., 2010. Aspects of neuroticism and the amygdala: chronic tuning from motivational styles. Neuropsychologia 48, 3399-3404.

Davidson, R.J., 1992. Anterior cerebral asymmetry and the nature of emotion. Brain Cogn. 20, 125-151.

Davidson, R.J., 1998. Affective style and affective disorders: perspectives from affective neuroscience. Cogn. Emotion 12, 307-330.

Davidson, R.J., Ekman, P., Saron, C.D., Senulis, J.A., Friesen, W.V., 1990. Approach-withdrawal and cerebral asymmetry: emotional expression and brain physiology. J. Person. Soc. Psychol. 58, 330-341.

Davidson, R.J., Marshall, J.R., Tomarken, A.J., Henriques, J.B., 2000. While a phobic waits: regional brain electrical and autonomic activity in social phobics during anticipation of public speaking. Biol. Psychiatry 47, 85-95.

deCharms, R.C., 2008. Applications of real-time fMRI. Nat. Rev. Neurosci. 9, 720-729.

Dehaene, S., Piazza, M., Pinel, P., Cohen, L., 2003. Three parietal circuits for number processing. Cogn. Neuropsychol. 20, 487–506.

De Pascalis, V., Cozzuto, G., Caprara, G.V., Alessandri, G., 2013. Relations among EEG-alpha asymmetry, BIS/BAS, and dispositional optimism. Biol. Psychol. 94, 198-209.

DeRubeis, R.J., Hollon, S.D., Amsterdam, J.D., Shelton, R.C., Young, P.R., Salomon, R.M., O'Reardon, J.P., Lovett, M.L., Gladis, M.M., Brown, L.L., Gallop, R., 2005. Cognitive therapy vs medications in the treatment of moderate to severe depression. Arch. Gen. Psychiatry 62, 409-416.

Dimidjian, S., Hollon, S.D., Dobson, K.S., Schmaling, K.B., Kohlenberg, R.J., Addis, M.E., Gallop, R., McGlinchey, J.B., Markley, D.K., Gollan, J.K., Atkins, D.C., Dunner, D.L., Jacobson, N.S., 2006. Randomized trial of behavioral activation, cognitive therapy, and antidepressant medication in the acute treatment of adults with major depression. J. Consult. Clin. Psychol. 74, 658-670.

Driessen, E., Hollon, S.D., 2010. Cognitive behavioral therapy for mood disorders: efficacy, moderators and mediators. Psychiatr. Clin. North Am. 33, 537-555.

Elliot, A.J., Covington, M.V., 2001. Approach and avoidance motivation. Edu. Psychol. Rev. 13, 73-92.

Ertl, M., Hildebrandt, M., Ourina, K., Leicht, G., Mulert, C., 2013. Emotion regulation by cognitive reappraisal – the role of frontal theta oscillations. NeuroImage 81, 412-421.

Friston, K.J., Buechel, C., Fink, G.R., Morris, J., Rolls, E., Dolan, R.J., 1997. Psychophysiological and modulatory interactions in neuroimaging. NeuroImage 6, 218-229.

Gilroy, S.W., Porteous, J., Charles, F., Cavazza, M., Soreq, E., Raz, G., Ikar, L., Or-Borichov, A., Ben-Arie, U., Klovatch, I., Hendler, T., 2013. A brain-computer interface to a plan-based narrative. Proc. 23$^{rd}$ Int. Joint Conf. Artif. Intel. (IJCAI 2013), 1997-2005.

Glover, G.H., Li, T.Q., Ress, D., 2000. Image-based method for retrospective correction of physiological motion effects in fMRI: RETROICOR. Magn. Reson. Med. 44, 162-167.

Gotlib, I.H., Ranganathan, C., Rosenfeld, J.P., 1998. Frontal EEG alpha asymmetry, depression, and cognitive functioning. Cogn. Emotion 12, 449-478.

Gruzelier, J.H., 2014. EEG-neurofeedback for optimizing performance. III: A review of methodological and theoretical considerations. Neurosci. Biobehav. Rev. 44, 159-182.

Hagemann, D., Naumann, E., Thayer, J.F., 2001. The quest for the EEG reference revisited: a glance from brain asymmetry research. Psychophysiol. 38, 847-857.

Hamilton, M., 1959. The assessment of anxiety states by rating. Br. J. Med. Psychol. 32, 50-55.

Hamilton, M., 1960. A rating scale for depression. J. Neurol. Neurosurg. Psychiatry 23, 56-62.

Harmon-Jones, E., Gable, P.A., Peterson, C.K., 2010. The role of asymmetric frontal cortical activity in emotion-related phenomena: a review and update. Biol. Psychol. 84, 451-462.

Heilman, K.M., 1996. Attentional asymmetries. In: Davidson, R.J., Hugdahl, K. (Eds.), Brain Asymmetry. The MIT Press, Cambridge, MA. 217-234.

Henriques, J.B., Davidson, R.J., 1991. Left frontal hypoactivation in depression. J. Abnorm. Psychol. 100, 535-545.

Herrington, J.D., Mohanty, A., Koven, N.S., Fisher, J.E., Stewart, J.L., Banich, M.T., Webb, A.G., Miller, G.A., Heller, W., 2005. Emotion-modulated performance and activity in left dorsolateral prefrontal cortex. Emotion 5, 200-207.

Herrington, J.D., Heller, W., Mohanty, A., Engels, A.S., Banich, M.T., Webb, A.G., Miller, G.A., 2010. Localization of asymmetric brain function in emotion and depression. Psychophysiol. 47, 442-454.

Johnston, S.J., Boehm, S.G., Healy, D., Goebel, R., Linden, D.E.J., 2010. Neurofeedback: a promising tool for the self-regulation of emotion networks. NeuroImage 49, 1066-1072.

Johnston, S., Linden, D.E.J., Healy, D., Goebel, R., Habes, I., Boehm, S.G., 2011. Upregulation of emotional areas through neurofeedback with a focus on positive mood. Cogn. Affect. Behav. Neurosci. 11, 44-51.

Kalpouzos, G., Nyberg, L., 2010. Hemispheric asymmetry of memory. In: Hugdahl, K., Westerhausen, R. (Eds.), The Two Halves of the Brain: Information Processing in the Cerebral Hemispheres. The MIT Press, Cambridge, MA. 499-530.

Keune, P.M., Bostanov, V., Hautzinger, M., Kotchoubey, B., 2013. Approaching dysphoric mood: state-effects of mindfulness meditation on frontal brain asymmetry. Biol. Psychol. 93, 105-113.

Kinreich, S., Podlipsky, I., Jamshy, S., Intrator, N., Hendler, T., 2014. Neural dynamics necessary and sufficient for transition into pre-sleep induced by EEG neurofeedback. NeuroImage 97, 19-28.

Koush, Y., Rosa, M.J., Robineau, F., Heinen, K., Rieger, S.W., Weiskopf, N., Vuilleumier, P., Van De Ville, D., Scharnowski, F., 2013. Connectivity-based neurofeedback: dynamic causal modeling for real-time fMRI. NeuroImage 81, 422-430.

Lewis, R.S., Weekes, N.Y., Wang, T.H., 2007. The effect of a naturalistic stressor on frontal EEG asymmetry, stress, and health. Biol. Psychol. 75, 239-247.

Linden, D.E.J., Habes, I., Johnston, S.J., Linden, S., Tatineni, R., Subramanian, L., Sorger, B., Healy, D., Goebel, R., 2012. Real-time self-regulation of emotion networks in patients with depression. PLoS ONE 7, e38115 (1-10).

McMenamin, B.W., Shackman, A.J., Maxwell, J.S., Bachhuber, D.R.W., Koppenhaver, A.M., Greischar, L.L., Davidson, R.J., 2010. Validation of ICA-based myogenic artifact correction for scalp and source-localized EEG. NeuroImage 49, 2416-2432.

McNair, D.M., Lorr, M., Droppleman, L.F., 1971. Profile of Mood States. Educational and Industrial Testing Service, San Diego.

Meir-Hasson, Y., Kinreich, S., Podlipsky, I., Hendler, I., Intrator, N., 2014. An EEG finger-print of fMRI deep regional activation. NeuroImage 102, 128-141.

Montgomery, S.A., Asberg, M., 1979. A new depression scale designed to be sensitive to change. Br. J. Psychiatry 134, 382-389.

Mulert, C., Lemieux, L. (Eds.), 2010. EEG-fMRI: Physiological Basis, Technique, and Applications. Springer-Verlag, Berlin Heidelberg.

Murray, E.A., Wise, S.P., Drevets, W.C., 2011. Localization of dysfunction in major depressive disorder: prefrontal cortex and amygdala. Biol. Psychiatry 69, e43-e54.

Paquette, V., Beauregard, M., Beaulieu-Prevost, D., 2009. Effect of a psychoneurotherapy on brain electromagnetic tomography in individuals with major depressive disorder. Psychiatry Res. Neuroimaging 174, 231-239.

Peeters, F., Ronner, J., Bodar, L., van Os, J., Lousberg, R., 2014a. Validation of a neurofeedback paradigm: manipulating frontal EEG alpha-activity and its impact on mood. Int. J. Psychophysiol. 93, 116-120.

Peeters, F., Oehlen, M., Ronner, J., van Os, J., Lousberg, R., 2014b. Neurofeedback as a treatment for major depressive disorder – a pilot study. PLoS ONE 9, e91837 (1-6).

Pizzagalli, D.A., Nitschke, J.B., Oakes, T.R., Hendrick, A.M., Horras, K.A., Larson, C.L., Abercrombie, H.C., Schaefer, S.M., Koger, J.V., Benca, R.M., Pascual-Marqui, R.D, Davidson, R.J., 2002. Brain electrical tomography in depression: the importance of symptom severity, anxiety, and melancholic features. Biol. Psychiatry 52, 73-85.

Pizzagalli, D.A., Sherwood, R.J., Henriques, J.B., Davidson, R.J., 2005. Frontal brain asymmetry and reward responsiveness: a source-localization study. Psychol. Sci. 16, 805-813.

Price, J.L., Drevets, W.C., 2010. Neurocircuitry of mood disorders. Neuropsychopharmacology 35, 192-216.

Robineau, F., Rieger, S.W., Mermoud, C., Pichon, S., Koush, Y., Van De Ville, D., Vuilleumier, P., Scharnowski, F., 2014. Self-regulation of inter-hemispheric visual cortex balance through real-time fMRI neurofeedback training. NeuroImage 100, 1-14.

Rosenfeld, J.P., Cha, G., Blair, T., Gotlib, I.H., 1995. Operant (biofeedback) control of left-right frontal alpha power differences: potential neurotherapy for affective disorders. Biofeed. Self-Regul. 20, 241-258.

Ruiz, S., Lee, S., Soekadar, S.R., Caria, A., Veit, R., Kircher, T., Birbaumer, N., Sitaram, R., 2013. Acquired self-control of insula cortex modulates emotion recognition and brain network connectivity in schizophrenia. Hum. Brain Mapping 34, 200-212.

Schlund, M.W., Cataldo, M.F., 2010. Amygdala involvement in human avoidance, escape, and approach behavior. NeuroImage 53, 769-776.

Sergerie, K., Chochol, C., Armony, J.L., 2008. The role of the amygdala in emotional processing: a quantitative meta-analysis of functional neuroimaging studies. Neurosci. Biobehav. Rev. 32, 811–830.

Shtark, M.B., Verevkin, E.G., Kozlova, L.I., Mazhirina, K.G., Pokrovskii, M.A., Petrovskii, E.D., Savelov, A.A., Starostin, A.S., Yarosh, S.V., 2015. Synergetic fMRI-EEG brain mapping in alpha-rhythm voluntary control mode. Bull. Exp. Biol. Med. 158, 644-649.

Snaith, R.P., Hamilton, M., Morley, S., Humayan, A., Hargreaves, D., Trigwell, P., 1995. A scale for the assessment of hedonic tone: the Snaith-Hamilton Pleasure Scale. Br. J. Psychiatry 167, 99-103.

Spielberg, J.M., Miller, G.A., Engels, A.S., Herrington, J.D., Sutton, B.P., Banich, M.T., Heller, W., 2011. Trait approach and avoidance motivation: lateralized neural activity associated with executive function. NeuroImage 54, 661-670.

Spielberg, J.M., Miller, G.A., Warren, S.L., Engels, A.S., Crocker, L.D., Banich, M.T., Sutton, B.P., Heller, W., 2012. A brain network instantiating approach and avoidance motivation. Psychophysiol. 49, 1200-1214.

Spielberger, C.D., Gorsuch, R.L., Lushene, R.E., 1970. Test manual for the State Trait Anxiety Inventory. Consulting Psychologists Press, Palo Alto.

Stewart, J.L. Coan, J.A., Towers, D.N., Allen, J.J.B., 2011. Frontal EEG asymmetry during emotional challenge differentiates individuals with and without lifetime major depressive disorder. J. Affect. Disord. 129, 167-174.

Sulzer, J., Haller, S., Scharnowski, F., Weiskopf, N., Birbaumer, N., Blefari, M.L., Bruehl, A.B., Cohen, L.G., deCharms, R.C., Gassert, R., Goebel, R., Herwig, U., LaConte, S., Linden, D., Luft, A., Seifritz, E., Sitaram, R., 2013. Real-time fMRI neurofeedback: progress and challenges. NeuroImage 76, 386-399.

Sutton, S.K., Davidson, R.J., 1997. Prefrontal brain asymmetry: a biological substrate of the behavioral approach and inhibition systems. Psychol. Sci. 8, 204-210.

Talairach, J., Tournoux, P., 1988. Co-Planar Stereotaxic Atlas of the Human Brain. Thieme Medical Publishers, New York.

Thibodeau, R., Jorgensen, R.S., Kim, S., 2006. Depression, anxiety, and resting frontal EEG asymmetry: a meta-analytic review. J. Abnorm. Psychol. 115, 715-729.

Tomarken, A.J., Keener, A.D., 1998. Frontal brain asymmetry and depression: a self-regulatory perspective. Cogn. Emotion 12, 387-420.

Weiskopf, N., 2012. Real-time fMRI and its application to neurofeedback. NeuroImage 62, 682-692.

Young, K.D., Zotev, V., Phillips, R., Misaki, M., Yuan, H., Drevets, W.C., Bodurka, J., 2014. Real-time fMRI neurofeedback training of amygdala activity in patients with major depressive disorder. PLoS ONE 9, e88785 (1-13).

Zich, C., Debener, S., Kranczioch, C., Bleichner, M.G., Gutberlet, I., De Vos, M., 2015. Real-time EEG feedback during simultaneous EEG-fMRI identifies the cortical signature of motor imagery. NeuroImage 114, 438-447.

Zotev, V., Krueger, F., Phillips, R., Alvarez, R.P., Simmons, W.K., Bellgowan, P., Drevets, W.C., Bodurka, J., 2011. Self-regulation of amygdala activation using real-time fMRI neurofeedback. PLoS ONE 6, e24522 (1-17).

Zotev, V., Yuan, H., Phillips, R., Bodurka, J., 2012. EEG-assisted retrospective motion correction for fMRI: E-REMCOR. NeuroImage 63, 698-712.

Zotev, V., Phillips, R., Young, K.D., Drevets, W.C., Bodurka, J., 2013a. Prefrontal control of the amygdala during real-time fMRI neurofeedback training of emotion regulation. PLoS ONE 8, e79184 (1-14).

Zotev, V., Yuan, H., Misaki, M., Phillips, R., Young, K.D., Bodurka, J., 2013b. Real-time fMRI neurofeedback training of amygdala modulates frontal EEG asymmetry in MDD patients. Proc. Int. Soc. Magn. Reson. Med. (ISMRM) 21, 0521.

Zotev, V., Yuan, H., Misaki, M., Phillips, R., Young, K.D., Bodurka, J., 2013c. Real-time fMRI neurofeedback training of amygdala modulates frontal EEG asymmetry in MDD patients. Proc. Org. Hum. Brain Mapping (OHBM) 19, 3197.

Zotev, V., Phillips, R., Yuan, H., Misaki, M., Bodurka, J., 2014. Self-regulation of human brain activity using simultaneous real-time fMRI and EEG neurofeedback. NeuroImage 85, 985-995.

Zotev, V., Young, K.D., Phillips, R., Yuan, H., Misaki, M., Bodurka, J., 2015. Effects of real-time fMRI neurofeedback of the amygdala specific to major depressive disorder. Proc. Int. Soc. Magn. Reson. Med. (ISMRM) 23, 0512.

**Table 1. Participants' emotional state measures before and after the rtfMRI-nf session.** The emotional states were assessed using the Profile of Mood States (POMS) and the Visual Analogue Scale (VAS).

| Measure | Experimental group ($n$=13) | | | Control group ($n$=11) | | |
|---|---|---|---|---|---|---|
| | Before mean | After mean | Change $t$-score [$q$] | Before mean | After mean | Change $t$-score [$q$] |
| **POMS** | | | | | | |
| Depression | 14.8 | 8.77 | −3.03 [0.030]* | 10.0 | 9.09 | −0.69 [0.623] |
| Total mood disturbance | 38.5 | 22.5 | −2.34 [0.041]* | 32.3 | 28.6 | −0.51 [0.623] |
| **VAS** | | | | | | |
| Happiness | 3.92 | 5.38 | +2.30 [0.041]* | 4.18 | 5.27 | +1.54 [0.462] |

* FDR $q$<0.05

**Table 2. Group statistical data for the psychophysiological interaction (PPI) effect for frontal EEG asymmetry.** The PPI interaction [FEA-based regressor] × [Happy–Count] was defined using time-dependent frontal EEG asymmetry (FEA) for channels F3 and F4 ('F4–F3'). Location of the point with the peak group $t$-score and the number of voxels are specified for each cluster obtained after FWE correction for multiple comparisons.

| Region | Laterality | x, y, z (mm) | t-score | Size (vox.) |
|---|---|---|---|---|
| **Frontal Lobe** | | | | |
| Lateral orbitofrontal cortex (BA 47) | L | −33, 23, −6 | 9.25 | 239 |
| Medial frontal polar cortex (BA 10) | L | −13, 61, 20 | 7.64 | 200 |
| Medial frontal polar cortex (BA 9) | R | 3, 45, 20 | 7.94 | 148 |
| Superior frontal gyrus (BA 8) | L | −19, 25, 46 | 8.35 | 114 |
| Lateral orbitofrontal cortex (BA 47) | R | 31, 13, −18 | 6.93 | 37 |
| Lateral orbitofrontal cortex (BA 47) | L | −33, 9, −16 | 5.89 | 36 |
| Middle frontal gyrus (BA 8) | R | 23, 29, 38 | 6.01 | 28 |
| **Temporal Lobe** | | | | |
| Middle temporal gyrus (BA 39) | L | −53, −63, 22 | 6.19 | 95 |
| Middle temporal gyrus (BA 21) | L | −57, −3, −14 | 6.29 | 90 |
| Middle temporal gyrus (BA 21) | R | 41, 3, −24 | 6.11 | 42 |
| Superior temporal gyrus (BA38) | R | 51, 1 −6 | 7.27 | 42 |
| Middle temporal gyrus (BA 39) | R | 47, −63, 20 | 5.46 | 28 |
| **Limbic Lobe** | | | | |
| Posterior cingulate cortex (BA 30) | R | 7, −41, 2 | 8.91 | 1253 |
| Anterior cingulate cortex (BA 32/24) | L | −10, 43, 0 | 8.85 | 415 |
| Amygdala / Parahippocampal gyrus | L | −17, −3, −16 | 9.31 | 244 |
| Hippocampus | R | 31, −11, −16 | 7.87 | 52 |
| Parahippocampal gyrus (BA 36) | R | 27, −27, −12 | 6.94 | 39 |
| Amygdala | R | 19, −1, −18 | 7.47 | 24 |
| **Parietal Lobe** | | | | |
| Precuneus (BA 31) | R | 1, −69, 28 | 8.87 | 226 |
| **Occipital Lobe** | | | | |
| Lingual gyrus (BA 18) | L | −7, −79, −8 | 7.12 | 40 |
| **Sub-lobar Regions** | | | | |
| Thalamus, mediodorsal | R | 5, −21, 6 | 8.78 | 169 |
| Insula (BA 13) | L | −35, −21, 18 | 7.58 | 92 |
| Insula (BA 13) | L | −41, −17, 8 | 9.36 | 29 |

BA – Brodmann areas; L – left; R – right; $x, y, z$ – Talairach coordinates;
FWE corrected $p<0.05$ (Size – cluster size, minimum 24 voxels for uncorr. $p<0.001$).

**Table 3. Group statistical data for the psychophysiological interaction (PPI) effect for left frontal upper alpha EEG power.** The PPI interaction [EEG-power-based regressor] × [Happy–Count] was defined using time-dependent average of normalized upper alpha EEG powers for channels F3, F7, and FC5 positioned over the left prefrontal cortex. Location of the point with the peak group $t$-score and the number of voxels are specified for each cluster obtained after FWE correction for multiple comparisons.

| Region | Laterality | x, y, z (mm) | t-score | Size (vox.) |
|---|---|---|---|---|
| Medial/superior frontal gyrus (BA 8) | L | –11, 29, 40 | –7.08 | 75 |
| Anterior cingulate cortex (BA 32/24) | L | –9, 37, 2 | –6.53 | 99 |
| Globus pallidus, medial | R | 15, –3, –4 | –4.61 | 113 |
|    Amygdala (same cluster) | L | –17, –7, –11 | –3.70 | |
|    Amygdala (same cluster) | R | 21, –3, –10 | –3.84 | |
|    Hypothalamus (same cluster) | R | 2, –1, –9 | –4.59 | |

BA – Brodmann areas;  L – left;  R – right;  $x$, $y$, $z$ – Talairach coordinates;
FWE corrected $p<0.05$ (Size – cluster size, minimum 75 voxels for uncorr. $p<0.01$).

# Supplementary material

## S1.1. fMRI data analysis

Pre-processing of single-subject fMRI data included correction of cardiorespiratory artifacts using the AFNI implementation of the RETROICOR method (Glover et al., 2000). Further fMRI pre-processing involved slice timing correction and volume registration for all EPI volumes acquired in the experiment using the 3dvolreg AFNI program with two-pass registration. The fMRI data and motion parameters were Fourier low-pass filtered at 0.1 Hz. Standard general linear model (GLM) analysis was then performed for each of the seven fMRI runs (Fig. 1B) using the 3dDeconvolve AFNI program. The GLM model included two block-stimulus condition terms, Happy Memories and Count (Fig. 1B), represented by the standard block-design regressors in AFNI. A general linear test term was included to compute the Happy Memories vs Count contrast. Nuisance covariates included the six fMRI motion parameters and five polynomial terms for modeling the baseline. GLM $\beta$ coefficients were computed for each voxel, and average percent signal changes for Happy vs Rest, Count vs Rest, and Happy vs Count contrasts were obtained by dividing the corresponding $\beta$ values ($\times 100\%$) by the $\beta$ value for the constant baseline term. The resulting fMRI percent signal change maps for each run were transformed to the Talairach space using each subject's high-resolution anatomical brain image.

Average BOLD activity levels for the left and right amygdala were computed in the offline analysis for the LA and RA ROIs, exhibited in Fig. 2A. The ROIs were defined anatomically as the amygdala regions specified in the AFNI implementation of the Talairach-Tournoux atlas (TT_N27). The voxel-wise fMRI percent signal change data from the GLM analysis, transformed to the Talairach space, were averaged within the LA and RA ROIs and used as GLM-based measures of the amygdala BOLD activity.

To directly examine time courses of the amygdala BOLD activity, we first transformed the LA and RA ROIs (Fig. 2A) to each subject's individual high-resolution anatomical image space, and then to the individual EPI image space (all EPI volumes in the experiment were spatially registered together). The LA and RA ROIs in the EPI space included approximately 100 voxels each. In addition, bilateral 10-mm-diameter ROIs were defined within white matter (WM) and ventricle cerebrospinal fluid (CSF) and similarly transformed. The resulting ROIs in the individual EPI space were used as masks to obtain average time courses for the LA, RA, WM, and CSF regions. We then used the 3dDetrend AFNI program to orthogonalize the LA and RA time courses with respect to time courses of the polynomial-modeled baseline, the WM and CSF ROIs, the six fMRI motion parameters, and the Happy Memories and Count block-stimulus fMRI regressors.

To characterize fMRI signal-to-noise performance, we examined temporal signal-to-noise ratio (TSNR) of fMRI data for each participant. The TSNR is defined for each fMRI voxel as the ratio of mean fMRI signal and its temporal standard deviation, $TSNR=\mathrm{mean}(S(t))/\mathrm{std}(S(t))$. A TSNR map was computed for each subject's fMRI data for the initial Rest (RE) fMRI run after the RETROICOR correction and volume registration (without temporal filtering or spatial smoothing). The TSNR performance is illustrated for a representative subject in Fig. S1A, and for the EG group in Fig. S1B.

## S1.2. Amygdala BOLD laterality

To compare BOLD activity levels for the LA and RA, we considered amygdala BOLD laterality, i.e. differences in the mean GLM-based fMRI percent signal changes between the LA and RA ROIs for each task, run, and participant. Because the baseline fMRI signal levels, used in computation of percent signal changes, were somewhat different for the LA and RA, we also examined differences in their percent signal changes with respect to the average baseline, (baseline(LA)+baseline(RA))/2, as well as differences in mean GLM $\beta$ coefficients for the LA and RA ROIs without converting them to percent signal changes. The results of all three approaches were very similar in all group analyses involving the amygdala BOLD laterality. Therefore, we only report the differences in fMRI percent signal changes that were determined independently for the LA and RA.

The introduction of the amygdala BOLD laterality in the present study is justified for the following reasons. *First*, the two amygdala regions are quite compact and their respective mean BOLD activity levels can be determined fairly unambiguously (compared, for example, to extended networks of cortical regions involved in emotion regulation). *Second*, both the LA and the RA can exhibit fMRI activations that are not directly related to particular rtfMRI-nf properties. Such activations may reflect the overall positive emotion induction, as well as other changes in emotional state. *Third*, due to the relative proximity of the LA and RA, their apparent BOLD activity levels may be affected by similar fMRI artifacts. Therefore, the amygdala BOLD laterality in our study may conceivably provide a more sensitive measure of *target-specific* rtfMRI-nf effects.

## S1.3. EEG data analysis

Removal of MR and cardioballistic (CB) artifacts was based on the average artifact subtraction method implemented in BrainVision Analyzer 2. The MR artifact template was defined using MRI slice markers recorded with the EEG data. After the MR artifact removal, the EEG data were band-pass filtered between 0.5 and 80 Hz (48 dB/octave) and downsampled to 250 S/s sampling rate (4 ms interval). The fMRI slice selection frequency (17 Hz) and its harmonics were removed by band-rejection filtering. The CB artifact template was determined from the cardiac waveform recorded by the ECG channel, and the CB artifact to be subtracted was defined, for each channel, by a moving average over 21 cardiac periods. Intervals with strong motion artifacts were not included in the CB correction.

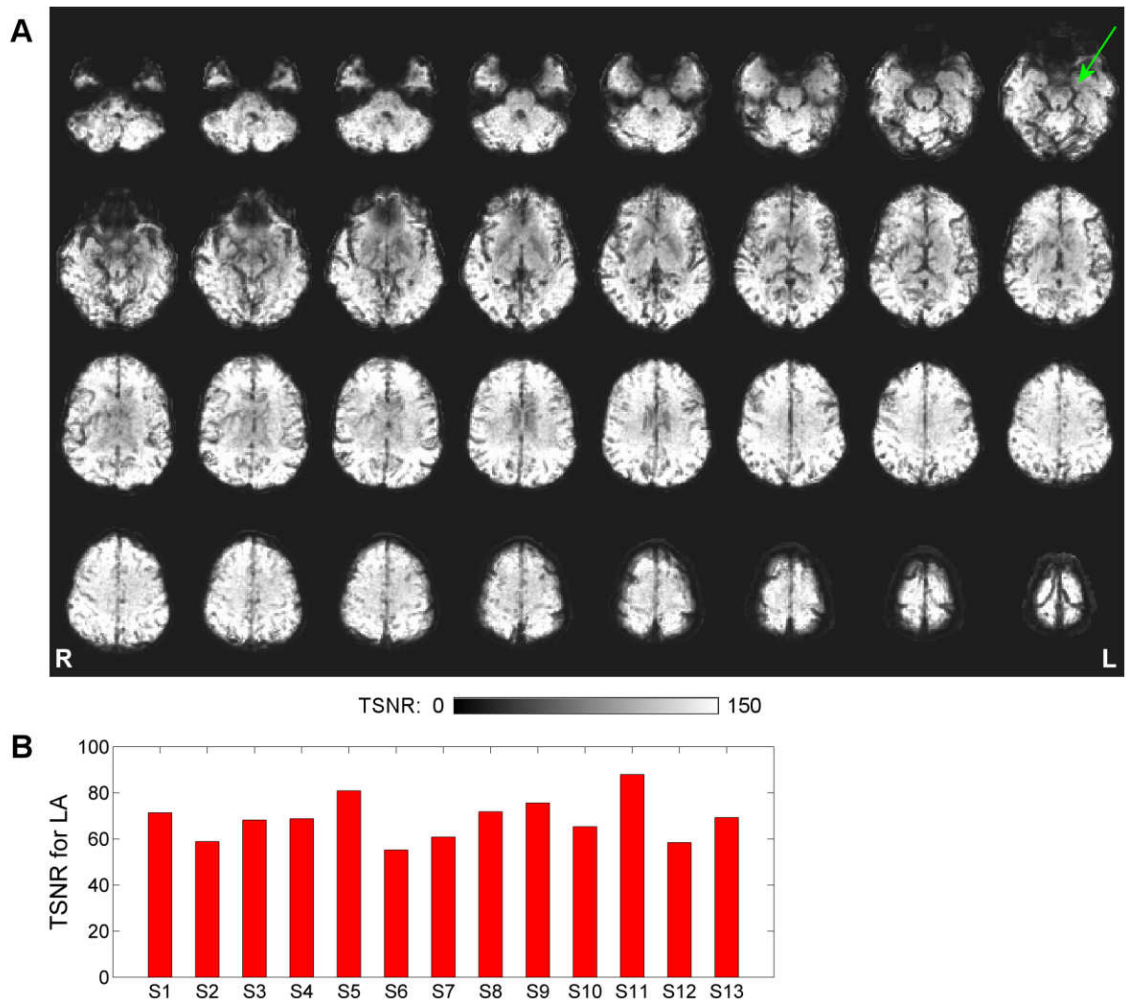

**Figure S1. fMRI signal-to-noise performance. A)** Temporal SNR (TSNR) map for a representative participant for the initial Rest fMRI run (Fig. 1B). The map shows 32 axial EPI slices with fMRI voxel size of $1.875 \times 1.875 \times 2.9$ mm$^3$. The green arrow points to the approximate center of the left amygdala (LA) ROI in the individual EPI space. The LA ROI was defined in the Talairach space (Fig. 2A) and transformed to each subject's individual EPI space as described in section S1.1. **B)** Average TSNR values for the LA ROI for 13 subjects (S1...S13) in the experimental group (EG) for the initial Rest fMRI run.

Following the MR and CB artifact removal, the EEG data from the seven experimental runs were concatenated to form a single dataset. The data were carefully examined, and intervals exhibiting significant motion or instrumental artifacts ("bad intervals") were excluded from the analysis.

Channel Cz was selected as a new reference, and FCz was restored as a regular channel. We did not use the average mastoid reference (e.g. Hagemann et al., 2001; Keune et al., 2013), because the mastoid EEG channels (TP9, TP10) exhibit strong motion-related artifacts inside an MRI scanner. These artifacts are caused by deformations of the mastoid electrode leads due to interactions with the soft padding underneath and on both sides of the head during head, jaw, and ear movements.

An independent component analysis (ICA) was performed over the entire dataset with exclusion of the bad intervals. This approach ensured that independent components (ICs) corresponding to various artifacts were identified and removed in a consistent manner across all seven runs. Channels TP9 and TP10 were excluded from the ICA and further analysis, because, as mentioned above, their signals are very sensitive to head, jaw, and ear movements, producing large artifacts. The Infomax ICA algorithm (Bell & Sejnowski, 1995), implemented in BrainVision Analyzer 2, was applied to the data from 29 EEG channels and yielded 29 ICs. Time courses, spectra, topographies, and kurtosis values of all the ICs were very carefully analyzed (see e.g. McMenamin et al., 2010 and supplement therein) to identify various artifacts, as well as EEG signals of neuronal origin, with particular attention to the alpha and theta EEG bands. After all the ICs had been classified, an inverse ICA transform was applied to remove the identified artifacts from the EEG data. Following the ICA-based artifact removal, the EEG data were low-pass filtered at 40 Hz (48 dB/octave) and downsampled to 125 S/s (8 ms interval). Because many artifacts had been already removed using the ICA, the data were examined again, and new bad intervals were defined to exclude remaining artifacts.

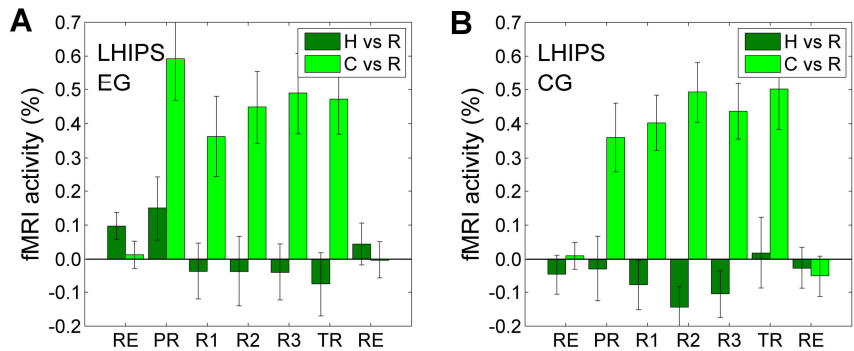

**Figure S2. BOLD activity levels for the LHIPS region during the rtfMRI-nf experiment. A)** Average fMRI percent signal changes for the left horizontal segment of the intraparietal sulcus (LHIPS) ROI (Fig. 1E) for the experimental group (EG). Each bar represents a mean GLM-based fMRI percent signal change for the LHIPS ROI with respect to the Rest baseline for the Happy Memories (H vs R, dark green) or Count (C vs R, green) conditions in a given run, averaged across the group. The error bars are standard errors of the means (sem). **B)** Average fMRI percent signal changes for the LHIPS ROI (Fig. 1E) for the control group (CG).

A time-frequency analysis was performed to compute EEG power for each channel as a function of time and frequency. The continuous wavelet transform with Morlet wavelets, implemented in BrainVision Analyzer 2, was applied to obtain EEG signal power in [0.25...15] Hz frequency range with 0.25 Hz frequency resolution and 8 ms temporal resolution. An average EEG power as a function of time was then computed for a frequency band of interest.

An EEG coherence analysis was conducted for the same data. It was applied to the Happy Memories conditions and performed separately for each of the four rtfMRI-nf runs. It included a segmentation with 4.096 s intervals, a complex FFT with 0.244 Hz spectral resolution, and the Coherence transform implemented in BrainVision Analyzer 2. A coherence value for signals from two EEG channels at a given frequency was computed as the squared magnitude of their cross spectrum value normalized by their power spectrum values at the same frequency ("cross-spectrum/ autospectrum"). An average coherence value for a frequency band of interest was then computed for each channel pair.

### S2.1. Self-report performance ratings

Self-report performance ratings provided information about each participant's success at recalling happy memories during a given run and happiness level immediately after that run. The two ratings (averaged for three rtfMRI-nf training runs) exhibited significant across-subjects correlations both for the EG (Hap vs Mem: $r=0.84$, $p<0.0003$) and the CG (Hap vs Mem: $r=0.68$, $p<0.022$). Remarkably, the EG and CG groups did not differ with respect to these self-report ratings. Independent-samples *t*-tests showed no significant group differences in either the happiness ratings (all $p>0.6$) or the memory-recall ratings (all $p>0.4$) for any of the runs. For example, the largest group difference in the happiness ratings was observed for Run 3 with mean happiness scores of 5.77 ($SD=2.04$) for the EG and 5.36 ($SD=1.67$) for the CG ($t(22)=0.50$, $p<0.619$).

### S2.2. Amygdala BOLD activity statistics

For the EG, the LA BOLD activity levels for the Happy Memories conditions (H vs R, Fig. 3A, *left*) were significant (after FDR correction for testing the five task runs) for the Transfer run (TR: $t(12)=4.64$, $q<0.005$) and showed a trend toward significance for the Practice run (PR: $t(12)=2.25$, $q<0.073$) and Run 3 (R3: $t(12)=2.62$, $q<0.055$). There was no significant LA activity difference for the Happy Memories conditions between the last rtfMRI-nf run and the Transfer run (TR vs R3: $t(12)=1.20$, $p<0.246$), indicating that the rtfMRI-nf training effect generalized beyond the actual training. For the CG, the LA BOLD activity levels for the Happy Memories conditions (H vs R, Fig. 3B, *left*) were not significant (after FDR correction for testing the five task runs), e.g. for Run 3 (R3: $t(10)=2.32$, $q<0.118$), and the Transfer run

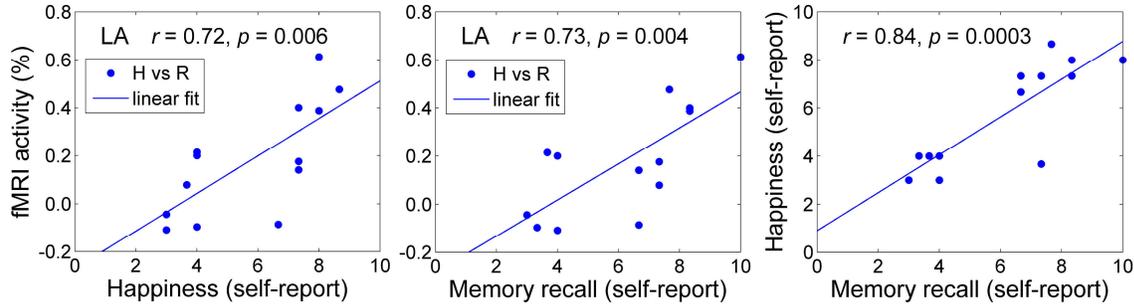

**Figure S3. Correlations between BOLD activity levels of the left amygdala during the rtfMRI-nf training and self-report performance ratings.** The results are for the experimental group (EG), with each data point corresponding to one subject. The participants provided self-report ratings for the happiness level and the success in recalling happy autobiographical memories after each run. Mean BOLD activity levels for the Happy Memories conditions with respect to the Rest baseline (H vs R) for the left amygdala (LA) were averaged across three rtfMRI-nf training runs (Run 1, Run 2, Run 3) for each subject. The self-report ratings were similarly averaged for the three rtfMRI-nf training runs. The plot on the right shows a strong correlation between the memory recall and happiness ratings.

(TR: $t(10)=2.26$, $q<0.118$).

The LA BOLD activity levels for the Happy Memories conditions for the EG (H vs R, Fig. 3A, *left*) exhibited a positive linear trend ('LT') that was significant across six experimental runs (LT(RE...TR): $F(1,12)=9.38$, $p<0.010$) and approached significance across five runs (LT(RE...R3): $F(1,12)=3.61$, $p<0.082$). In contrast, the LA BOLD activity levels for the CG (H vs R, Fig. 3B, *left*) did not exhibit a linear trend (LT(RE...TR): $F(1,10)=2.37$, $p<0.155$; LT(RE...R3): $F(1,10)=1.78$, $p<0.211$).

A two-way 4 (Training: PR, R1, R2, R3) × 2 (Group: EG, CG) repeated measures ANOVA applied to the Happy Memories BOLD activity levels for the LA (H vs R, Fig. 3, *left*) did not reveal significant effects. However, a similar 4 × 2 repeated measures ANOVA applied to the Happy Memories vs Count (H vs C) activity contrasts for the LA revealed a significant effect for the Group ($F(1,22)=4.96$, $p<0.036$).

Figure S2 illustrates fMRI activity of the LHIPS region (Fig. 1E). Notably, the average Happy Memories vs Rest BOLD activity levels for the LHIPS ROI were negative during the rtfMRI-nf runs for the CG (H vs R, dark green in Fig. S2B), despite the fact that the LHIPS was used as a target ROI for the rtfMRI-nf for the CG.

A two-way 5 (Training: PR, R1, R2, R3, TR) × 2 (ROI: LA, LHIPS) repeated measures ANOVA applied to the Happy Memories BOLD activity levels for the LA and LHIPS ROIs for the EG showed a significant effect for ROI ($F(1,12)=4.95$, $p<0.046$), and a Training×ROI interaction effect that showed a trend toward significance ($F(4,48)=2.28$, $p<0.074$). For the CG, the ROI effect was also significant ($F(1,10)=17.2$, $p<0.002$), while the Training×ROI interaction effect was insignificant ($F(4,40)=0.93$, $p<0.456$).

*S2.3. Correlations between amygdala BOLD activity and self-report performance ratings*

Figure S3 illustrates correlations between average BOLD activity levels for the LA ROI and the average self-report performance ratings for the EG participants. The average individual Happy Memories vs Rest (H vs R) LA activity levels significantly correlated with the corresponding average happiness and average memory-recall ratings (Fig. S3, *left*, *middle*). Notably, these two ratings strongly correlated with each other (Fig. S3, *right*). For the RA ROI, correlations between the average BOLD activity levels and these ratings were also positive, but not significant (EG, RA vs Hap: $r=0.39$, $p<0.185$; RA vs Mem: $r=0.55$, $p<0.054$). For the CG, no correlations were found between the average LA activity levels and either the happiness ratings or the memory-recall ratings (CG, LA vs Hap: $r=-0.22$, $p<0.516$; LA vs Mem: $r=-0.13$, $p<0.710$).

Partial correlation analyses for the three quantities in Fig. S3 yielded the following results. The partial correlation between the LA activity and the happiness ratings while controlling for the memory-recall: $r(10)=0.28$, $p<0.385$. The partial correlation between the LA activity and the memory-recall ratings while controlling for the happiness: $r(10)=0.35$, $p<0.269$. Comparison with the zero-order correlations in Fig. S3 suggests that either of the two ratings (happiness or memory-recall) plays some mediating role when correlation between the LA activity and the other rating (i.e. memory-recall or happiness) is considered.

*S2.4. Correlations between amygdala BOLD activity and psychological measures*

Figure S4 illustrates correlations between average BOLD activity levels for the LA and RA ROIs and individual psychological measures for the EG participants. The average individual Happy Memories (H vs R) LA activity levels for the rtfMRI-nf training runs significantly correlated with state changes in happiness (Fig. S4, *top*

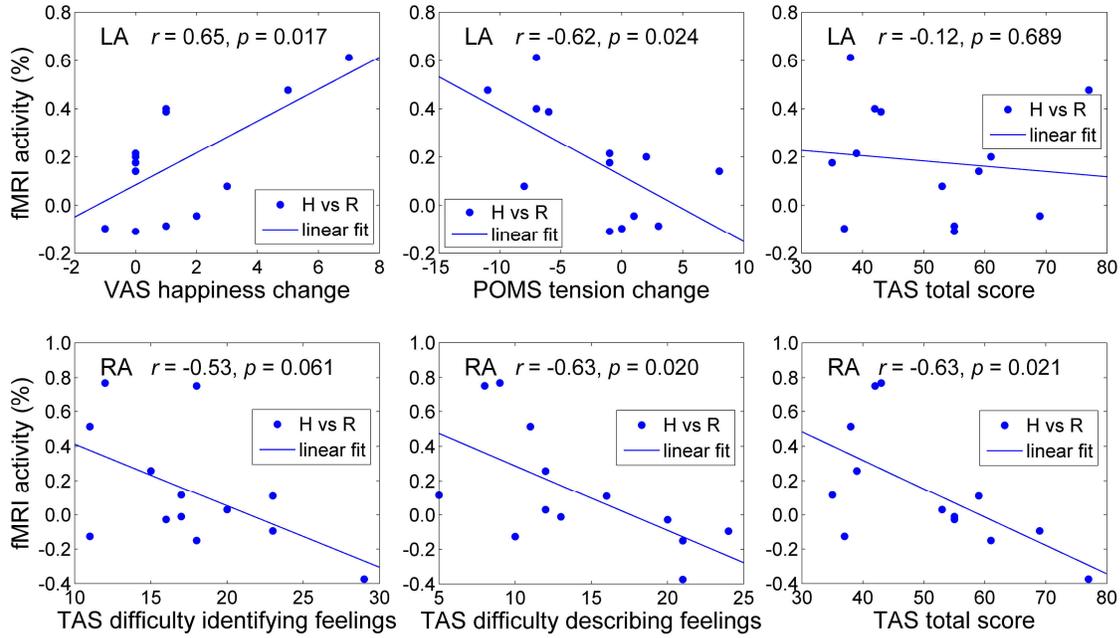

**Figure S4. Correlations between BOLD activity levels of the amygdala during the rtfMRI-nf training and individual psychological measures.** Each data point corresponds to one participant in the experimental group (EG). Mean BOLD activity levels for the Happy Memories conditions with respect to the Rest baseline (H vs R) for the left or right amygdala (LA, RA) were further averaged across three rtfMRI-nf training runs (Run 1, Run 2, Run 3) for each subject. Changes in emotional states were characterized by differences in psychological scores measured after and before the experiment. Abbreviations: VAS – Visual Analogue Scale, POMS – Profile of Mood States, TAS – Toronto Alexithymia Scale.

*left*) and inversely correlated with state changes in tension (Fig. S4, *top middle*). The average RA activity levels exhibited significant inverse correlations with trait measures of alexithymia (Fig. S4, *bottom middle, right*). Notably, these measures strongly correlated with each other (TAS difficulty describing feelings vs TAS total score: $r=0.88$, $p=0.00008$). The average activity levels for the LA also showed inverse correlations with the TAS measures, which, however, were not significant (LA vs TAS difficulty identifying feelings: $r=-0.01$, $p<0.978$; LA vs TAS difficulty describing feelings: $r=-0.23$, $p<0.452$; LA vs TAS total score: $r=-0.12$, $p<0.689$, Fig. S4, *top right*). Interestingly, for the CG, correlations with the TAS measures (not shown) were comparable for the LA and RA (CG, LA vs TAS difficulty describing feelings: $r=-0.58$, $p<0.062$; RA vs TAS difficulty describing feelings: $r=-0.53$, $p<0.097$).

Partial correlation analyses for the three quantities in Fig. S4 (*bottom middle, right*) provided the following results. The partial correlation between the RA activity and the TAS difficulty describing feelings while controlling for the TAS total score: $r(10)=-0.22$, $p<0.500$. The partial correlation between the RA activity and the TAS total score while controlling for the TAS difficulty describing feelings: $r(10)=-0.20$, $p<0.539$. Comparison with the zero-order correlations in Fig. S4 suggests that either of the two TAS scores may play some mediating role when correlation between the RA activity and the other score is examined.

### S2.5. Correlations between amygdala BOLD laterality and psychological measures

Figure S5 illustrates correlations between average amygdala BOLD laterality (denoted 'LA–RA') and psychological measures for the participants in the EG. The average amygdala laterality exhibited a positive and approaching significance correlation with the HDRS depression severity ratings (Fig. S5, *left*). No such correlations were observed for either the LA or the RA separately (LA vs HDRS: $r=0.34$, $p<0.263$; RA vs HDRS: $r=-0.26$, $p<0.384$). The average amygdala BOLD laterality also showed a significant positive correlation with the TAS total scores (Fig. S5, *middle*). It should be noted that trait alexithymia measures (TAS-20) positively correlated with individual HDRS depression severity ratings both for the EG (TAS difficulty in externally oriented thinking vs HDRS: $r=0.71$, $p<0.006$; TAS total score vs HDRS: $r=0.62$, $p<0.024$, Fig. S5, *right*) and for the CG (TAS difficulty identifying feelings vs HDRS: $r=0.83$, $p<0.002$).

Partial correlation analyses for the three quantities in Fig. S5 produced the following results. The partial correlation between the LA–RA and the HDRS ratings while controlling for the TAS total score: $r(10)=0.25$, $p<0.427$. The partial correlation between the LA−RA and the TAS total scores while controlling for the HDRS depression severity: $r(10)=0.43$, $p<0.165$. Comparison with the zero-order correlations in Fig. S5 suggests that the TAS total score may have a stronger mediating effect on the LA−RA vs HDRS correlation than the HDRS rating on the LA−RA vs TAS correlation. Indeed, a Sobel mediation test showed an approaching significance mediation effect ($z=1.913$, $p<0.056$) of the TAS total score on the correlation between the HDRS ratings and the amygdala BOLD laterality.

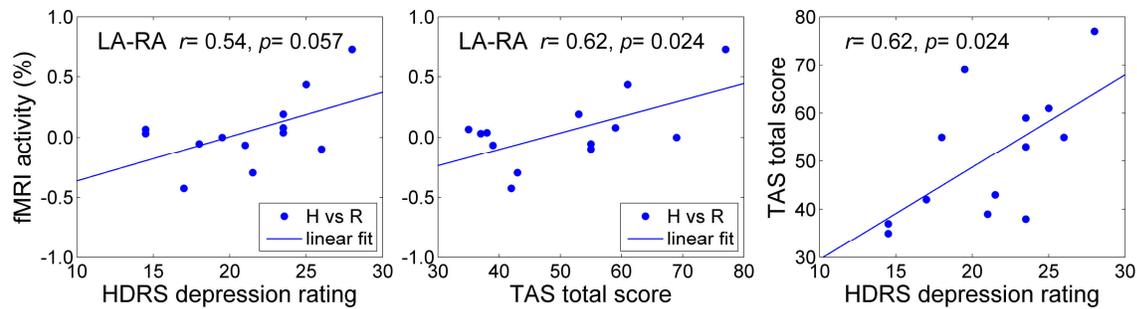

**Figure S5. Correlations between amygdala BOLD laterality during the rtfMRI-nf training and individual psychological measures.** The results are for the experimental group (EG), with each data point corresponding to one participant. Differences in mean BOLD activity levels for the Happy Memories conditions with respect to the Rest baseline (H vs R) between the left and right amygdala ('LA−RA') were further averaged across four rtfMRI-nf runs (Practice, Run 1, Run 2, Run 3) for each subject. The plot on the right shows a correlation between the HDRS and TAS ratings. Abbreviations: HDRS – Hamilton Depression Rating Scale, TAS – Toronto Alexithymia Scale.

### S3.1. Discussion of the rtfMRI-nf effects on the amygdala BOLD activity

The rtfMRI-nf training effects on the amygdala BOLD activity for the EG in the present study (Fig. 3A) were similar to those reported previously for neurofeedback-naive healthy participants (Zotev et al., 2011) and MDD patients (Young et al., 2014). Specifically, the LA BOLD activity during the Happy Memories conditions was either significant or showing a trend toward significance for most of the task runs (see S2.2). The LA activity for the EG also exhibited an approaching significance positive linear trend across rtfMRI-nf runs (S2.2). However, the rtfMRI-nf effects for the CG in the present work (Fig. 3B) were different from those observed previously. While in the previous studies the LA BOLD activity levels for the CG exhibited negative trends across experimental runs (Young et al., 2014; Zotev et al., 2011), such trend was positive, though not significant, in our present analysis (S2.2). We have the following tentative explanation for this change in performance. During the first rtfMRI-nf session, the neurofeedback-naive participants in the CG expected to be able to effectively control the rtfMRI-nf signal, and their apparent inability to do so (with the sham rtfMRI-nf) might have led to progressively increased confusion and frustration that impaired performance. During the second rtfMRI-nf session, described in the present paper, the same CG participants had lower performance expectations, and, consequently, experienced less frustration. Conceivably, they might have also paid less attention to the rtfMRI-nf signal and focused more on the positive emotion induction.

### S3.2. Discussion of correlations between amygdala BOLD activity and self-report ratings

The self-reported performance ratings, obtained after each experimental run, provided insights into the participants' emotional states during the rtfMRI-nf session. Notably, the ratings for success at recalling happy memories and for happiness level did not significantly differ between the EG and CG groups. The memory-recall and happiness ratings significantly correlated both for the EG and CG (see S2.1), indicating that successful recall of happy autobiographical memories was essential for inducing positive emotion in both groups. However, significant positive correlations between the average LA BOLD activity levels and these ratings were observed only for the EG (Fig. S3), and not for the CG (see S2.3). This finding suggests that the rtfMRI-nf targeting the LA ROI did not significantly alter the overall emotion induction process for the EG (compared to the CG in the present study), but only *modified* it to selectively increase BOLD activity of the LA.

### S3.3. Discussion of correlations between amygdala BOLD activity and psychological measures

The amygdala BOLD activity levels during the rtfMRI-nf task also correlated with individual psychological measures acquired before (or both before and after) the rtfMRI-nf session (see S2.4 and Fig. S4). The average individual LA activity levels for the EG showed a significant positive correlation with state changes in happiness (VAS, Fig. S3), which is consistent with the result in Fig. S3, and a significant inverse correlation with state changes in tension (POMS, Fig. S4). The average individual RA activity levels showed significant inverse correlations with trait measures of alexithymia (TAS-20, Fig. S4). For the LA, such correlations were also inverse but non-significant (see S2.4). We interpret this result as follows. Difficulties in identifying and describing feelings (TAS-20) can lead to less efficient positive emotion induction and reduced levels of BOLD activity for both the LA and RA. (Such reduction is also observed for the CG). However, the rtfMRI-nf procedure targeting the LA modifies and enhances the emotion induction process to selectively increase BOLD activity of the LA. This leads to a stronger correlation between the resulting state happiness levels and the LA activity levels. This rtfMRI-nf effect can also make the LA BOLD activity levels less dependent on the overall emotion induction (which affects activities of both the LA and the RA) and less correlated with the trait alexithymia measures.

*S3.4. Discussion of correlations between amygdala BOLD laterality and psychological measures*

The average amygdala BOLD laterality ('LA–RA') during the rtfMRI-nf task for the EG showed positive correlation with the HDRS depression severity ratings that approached significance (see S2.5 and Fig. S5, *left*). However, the laterality also exhibited significant positive correlation with the total alexithymia score (TAS-20, Fig. S5, *middle*), which reflected differences in correlations with this score between the LA and RA (Fig. S4), interpreted above. The total alexithymia score itself significantly correlated with the HDRS ratings (Fig. S5, *right*). The Sobel mediation test suggested that the correlation between the HDRS depression severity and the amygdala BOLD laterality may be mediated by the MDD patients' alexithymia.

**Table S1. Participants' psychological trait measures before the rtfMRI-nf session.** The psychological traits were assessed using the Hamilton Depression Rating Scale (HDRS), the Montgomery-Asberg Depression Rating Scale (MADRS), the Hamilton Anxiety Rating Scale (HARS), the Snaith-Hamilton Pleasure Scale (SHAPS), the State-Trait Anxiety Inventory (STAI), and the Toronto Alexithymia Scale (TAS-20).

| Measure | Experimental group, mean (SD) | Control group, mean (SD) | Difference *t*-score [*p*]# |
|---|---|---|---|
| **HDRS** | 20.5 (4.0) | 20.9 (3.3) | –0.29 [0.778] |
| **MADRS** | 27.4 (6.8) | 28.5 (3.0) | –0.50 [0.620] |
| **HARS** | 17.5 (4.7) | 19.3 (5.2) | –0.86 [0.401] |
| **SHAPS** | 30.4 (6.1) | 32.8 (6.7) | –0.89 [0.383] |
| **STAI** | | | |
| Anxiety (trait) | 54.0 (8.5) | 51.4 (10) | +0.66 [0.518] |
| **TAS-20** | | | |
| Diff. identifying feelings | 17.7 (5.0) | 18.7 (5.1) | –0.48 [0.635] |
| Diff. describing feelings | 14.0 (5.6) | 15.6 (4.4) | –0.75 [0.464] |
| Diff. externally oriented | 19.3 (4.3) | 19.3 (5.2) | +0.02 [0.986] |
| Total alexithymia score | 51.0 (13) | 53.6 (11) | –0.51 [0.615] |

# two-tailed, uncorr.